\documentclass{aa}

\usepackage{txfonts}
\usepackage{graphicx, color}
\usepackage[colorlinks]{hyperref}
\usepackage{amsmath}
\usepackage{pdflscape}
\usepackage{multirow,bigdelim}
\usepackage{calc}
\usepackage{threeparttable}

\usepackage[normalem]{ulem}

\usepackage{caption}
\usepackage[labelformat = empty,position=top]{subcaption}
\usepackage{mathptmx}

\usepackage{xcolor}
\usepackage{multirow}
\usepackage{dirtytalk}

\definecolor{dgreen}{rgb}{0., 0.7, 0.}

\newcommand{\feh}{\mbox{\rm [{\rm Fe}/{\rm H}]}}

\newcommand{\Msun}{\mbox{$\mathrm{M}_{\odot}$}}
\newcommand{\Lsun}{\mbox{$\mathrm{L}_{\odot}$}}
\newcommand{\Rsun}{\mbox{$\mathrm{R}_{\odot}$}}

\newcommand{\He}{\mbox{$^3\mathrm{He}$}}
\newcommand{\Li}{\mbox{$^7\mathrm{Li}$}}

\newcommand{\Teff}{\mbox{$\mathrm{T}_{\mathrm{eff}}$}}

\usepackage[T1]{fontenc}
\usepackage{ae,aecompl}
{}

\usepackage{fixltx2e}
{\left\lbrace\begin{array}{@{}l@{}}}%
{\end{array}\right.}

\begin{document}

\title{A combined study of thermohaline mixing and envelope overshooting with PARSEC: Calibration to NGC 6397 and M4}
\titlerunning{Thermohaline mixing with \textsc{parsec}}

\subtitle{}

   \author{C. T. Nguyen
        \inst{1,2}
        \and
          A. Bressan
        \inst{2}
        \and
          A.J. Korn %
        \inst{3}
        \and
          G. Cescutti %
        \inst{1}
        \and
          G. Costa %
        \inst{4}
        \and
          F. Addari
        \inst{2} 
        \and
          L. Girardi %
        \inst{5}
        \and
          X. Fu
        \inst{6}
        \and
          Y. Chen
        \inst{7}
        \and
          P. Marigo %
        \inst{8}
          }

        \institute{
        INAF Osservatorio Astronomico di Trieste, Via Giambattista Tiepolo, 11, Trieste, Italy,\\
        \email{chi.nguyen@inaf.it}
        \and
        SISSA, Via Bonomea 265, I-34136 Trieste, Italy,
        \and
        Division of Astronomy and Space Physics, Department of Physics and Astronomy, Uppsala University, Box 516, SE-75120 Uppsala, Sweden,
        \and
        Univ Lyon, Univ Lyon1, ENS de Lyon, CNRS, Centre de Recherche Astrophysique de Lyon UMR5574, F-69230 Saint-GenisLaval, France
        \and
        INAF Osservatorio Astronomico di Padova, Vicolo dell'Osservatorio n. 5, Padova, Italy
        \and
        Purple Mountain Observatory, Chinese Academy of Sciences, Nanjing 210023, China
        \and
        Anhui University, Hefei 230601, China
        \and
        Dipartimento di Fisica e Astronomia, Universit\`a degli studi di Padova,
        Vicolo Osservatorio 3, Padova, Italy,
      }

\authorrunning{Nguyen et al.}

   \date{}

\hypersetup{
    linkcolor=blue,
    citecolor=blue,
    filecolor=magenta,      
    urlcolor=blue
}


\abstract{Thermohaline mixing is one of the main processes in low-mass red giant stars that affect the transport of chemicals and, thus, the surface abundances along the evolution. 
The interplay of thermohaline mixing with other processes, such as the downward overshooting from the convective envelope, should be carefully investigated. 
This study aims to understand the combined effects of thermohaline mixing and envelope overshooting.
After implementing the thermohaline mixing process in the \textsc{parsec} stellar evolutionary code, we compute tracks and isochrones (with \textsc{trilegal} code) and compare them with observational data. 
To constrain the efficiencies of both processes, we perform a detailed modelling that is suitable for globular clusters NGC 6397 and M4.
Our results indicate that an envelope overshooting efficiency parameter, $\Lambda_\mathrm{e}=0.6$, and a thermohaline efficiency parameter, $\alpha_\mathrm{th}=50$, are necessary to reproduce the RGB bump magnitudes and lithium abundances observed in these clusters. 
We find that both envelope overshooting and thermohaline mixing have a significant impact on the variation of $^7$Li abundances. 
Additionally, we also explore the effects of adopting solar-scaled or $\alpha$-enhanced mixtures on our models.
The $^{12}$C and the $^{12}$C/$^{13}$C ratio are also effective indicators to probe extra mixing in RGB stars. Although, their usefulness is currently limited by the lack of precise and accurate C-isotopes abundances.}

   \keywords{Stars: evolution - Stars: abundances - Stars: low-mass - Stars: pre-main sequence.}
   
   \maketitle
   
\section{Introduction}

Globular Clusters (GCs) are known as formidable laboratories for studying the properties of stars in our Galaxy. 
These systems are populated by long-lived low-mass stars and their chemical composition varies from a very low metallicity to an almost solar one. They thus provide extremely useful  information about the formation and evolution of the chemical elements of their birth clouds \citep{2005ARA&A..43..531B,2015ARA&A..53..631F}.
Furthermore, they are so populous that even the fastest stellar evolutionary phases can be seen and studied. They are thus an ideal workbench to probe different physical processes that happen inside stars. One such process, whose understanding is of paramount importance to build realistic stellar models, is internal mixing.

Especially during the first-ascent red-giant branch (RGB), due to the penetration of the convective envelope to the inner layers, mixing occurs between these two regions and leads to changes in surface abundance. This process is the so-called the first-dredge-up (1DU). 
In classical models, after the 1DU is completed, the surface abundances remain constant until the end of the RGB phase \citep[see][]{1992MNRAS.256..449S,1998A&A...332..204C}.
However, measurements of the surface abundances of RGB stars show a different, more nuanced picture. For instance, \cite{2000A&A...354..169G} showed the variation of many elements of field stars with $-$2<[Fe/H]<$-$1, as well as the trend of lithium abundances in five globular clusters in \citet{2022A&A...657A..33A}. A striking example is the case of NGC 6397, where the lithium abundances have been shown a significant drop after the RGB bump \citep{2009A&A...503..545L}, as well as the carbon abundance \citep{1990ApJ...359..307B}. All measurements agree that extra mixing after the 1DU is required.

The process of thermohaline mixing has been studied in great detail by many authors. It was first introduced in stellar astrophysics by \citet{1972ApJ...172..165U} and \citet{1980A&A....91..175K}. Later,  \citet{2006Sci...314.1580E} investigated the instability regions inside low-mass stars and found a mean molecular weight inversion that is suitable for thermohaline mixing to occur naturally by the reaction $^3\text{He}(^3\text{He},2\text{p})^4\text{He}$ above the H-burning shell. \citet{2007A&A...467L..15C} established the importance of thermohaline mixing by showing the significant change in surface abundances of red giant stars after the bump. Since then, many investigations on the mixing scheme have been published. For instance, \citet{2009MNRAS.396.2313S} pursued the effect of thermohaline mixing in carbon-normal and carbon-rich metal-poor giant stars \citep[see also][]{2009MNRAS.394.1051S,2014ApJ...797...21P}; and \citet{2017MNRAS.469.4600H,2018ApJ...863L...5H} studied the modifications to the standard thermohaline mixing model. 
Other processes can affect the efficiency of thermohaline mixing, e.g.\ rotation \citep[]{2011A&A...536A..28L,2013A&A...553A...1M} or magnetic field inhibits the mixing in Ap stars \citep{2007A&A...476L..29C}.

Here, we want to investigate the interplay between convective envelope overshooting and the efficiency of thermohaline mixing, and how they impact the evolution of stars. 
For this purpose, we first implemented thermohaline mixing in the \textsc{parsec} stellar evolutionary code \citep{2022A&A...665A.126N}\footnote{\url{http://stev.oapd.inaf.it/PARSEC/}} and calculated many low-mass models (in the mass range $0.64\leq$ M/$\Msun\leq 0.94$ ) with four initial metallicities, Z$=0.0001,0.0002,0.001,0.002$ (in mass fraction), with variations of envelope overshooting and thermohaline mixing efficiencies, and adopting solar-scaled and $[\alpha/\mathrm{Fe}]=0.4$-enhanced mixtures. All models are computed from the beginning of the pre-main-sequence (PMS) to the end of the RGB phase. The evolution of several chemical isotopes was followed during the entire evolution of the stars. The conversion from stellar evolutionary tracks to isochrones is done by using the \textsc{trilegal} code \citep{2005A&A...436..895G,2017ApJ...835...77M}\footnote{\url{http://stev.oapd.inaf.it/cgi-bin/cmd}}. In which more chemical isotopes are included in the isochrone tables, such as Li$^7$, C$^{13}$ in order to investigate the effect of thermohaline mixing is done in this work.

The paper is organised into five sections. Sect. \ref{theoretical_description} shows the theoretical model of thermohaline mixing as a diffusive process and how we include it in our code, together with the input physics used in this work. The impact of thermohaline mixing on the evolution of the stellar model is presented here. We also discuss the role of the chemical compositions that are used in computing the models in this section.
Then, Sect. \ref{alpha_th_calib} presents the calibration of the thermohaline efficiency parameter in which the variation of predicted surface lithium abundance is directly compared to the observed data, and the calibration of envelope overshooting efficiency base on the relative luminosity distance between the Li-peak and the onset of thermohaline mixing. Sect. \ref{EOV_calib} shows an independent calibration for the envelope overshooting efficiency parameter on the GC M4 based on seismic data of RGB stars. Finally, we will summarise and conclude this paper in Sect. \ref{conclude}.

\section{PARSEC models with thermohaline mixing}\label{theoretical_description}
\subsection{Thermohaline mixing: formalism}

Thermohaline mixing sets in when an inversion of molecular weight ($\mu$) arises in the interiors of a star. The nuclear reaction of two $^3$He to create one $^4$He and two protons, $^3\mathrm{He}(^3\mathrm{He,2p})^4\mathrm{He}$, below the convective envelope is the main mechanism that causes the instability regions of thermohaline mixing. However, in this study, we notice that the inversion of $\mu$-profile might also be caused by the atomic ionisation throughout the stars. This can be understood by the temperature distribution within a star. Namely, towards inner layers the star has higher temperatures, hence directly affecting the ionisation degree of the chemical isotopes, hence directly affecting the variation of molecular weight.

The physical process responsible for thermohaline mixing is the double-diffusive instability which is verified by \citet{2007A&A...467L..15C}. While by using the 3D hydrodynamics simulation \citet{2006Sci...314.1580E} claims the Rayleigh-Taylor instability. 
In either case, the instability happens in a stratified medium that satisfies the Ledoux criterion for convective stability, namely
\begin{align}
    \nabla_\mathrm{ad} - \nabla + \left(\frac{\phi}{\delta}\right)\nabla_\mu > 0,
\end{align}
where the inversion in mean molecular weight, i.e., $\nabla_\mu<0$, also must occurs. In the formula above, $\nabla_\mu=\mathrm{d}\ln\mu/\mathrm{d}\ln P$ is the molecular weight gradient, $\nabla=\partial\ln T/\partial\ln P$ is the temperature gradient, $\phi=(\partial\ln\rho/\partial\ln\mu)_\mathrm{P,T}$ and $\delta~=~-~(\partial\ln\rho/\partial\ln T)_\mathrm{P,\mu}$ are the thermo-dynamical derivatives, and $\nabla_\mathrm{ad}=P\delta/T\rho c_P$ is the adiabatic temperature gradient.

The mixing produced by the thermohaline instability in radiative regions is described by the following diffusion coefficient \citep[see][]{2010A&A...521A...9C},
\begin{align}\label{Dth}
    D_\mathrm{th}=\frac{3}{2}\alpha_\mathrm{th}K\frac{-\frac{\phi}{\delta}\nabla_\mu}{\nabla_\mathrm{ad}-\nabla},\quad \nabla_\mu<0,
\end{align}
where, $K=4acT^3/3(c_P\kappa\rho^2)$ is the thermal diffusivity, where $a$ is the radiation density constant, $c$ is the speed of light, $c_P$ is the heat capacity at constant pressure, $\kappa$ is the Rosseland mean opacity, $(\rho, T)$ are stellar density and temperature; the dimensionless parameter $\alpha_\mathrm{th}$ is a free parameter which controls the efficiency of thermohaline mixing. Therefore, the calibration of $\alpha_\mathrm{th}$ is needed and it will be discussed in detail in Sect. \ref{alpha_th_calib}.
By comparing Eq. \ref{Dth} with similar ones found in literature we obtain the following relation
\begin{align}
    \alpha_\mathrm{th}=\frac{2}{3}C_\mathrm{t}=\frac{16}{9}(\pi^2\alpha^2),
\end{align}
where $C_\mathrm{t}$ and $\alpha$ are the thermohlaine efficiency parameters used in \citet{2007A&A...467L..15C} and \citet{1972ApJ...172..165U}, respectively.

\subsection{\textsc{PARSEC} models} \label{parsec_th_implication}
We implemented the thermohaline mixing process in the latest version, \textsc{parsec}~V2.0\citep{2019A&A...631A.128C,2019MNRAS.485.4641C,2022A&A...665A.126N}.
For seek of brevity, we remind that the standard \textsc{parsec}~V1.2S code is thoroughly described in \citep{2012MNRAS.427..127B,2014MNRAS.444.2525C,2018MNRAS.476..496F}. In this version  \textsc{parsec}~V2.0,
the nuclear reaction network includes the p-p chains, the CNO tri-cycle, the Ne-Na and Mg-Al chains, 
$^{12}$C, $^{16}$O and $^{20}$Ne burning reactions,
and the $\alpha$-capture reactions up to $^{56}$Ni, for a total of 72 different reactions tracing 32 isotopes: 
$^1$H, D, $^3$He, $^4$He,   $^7$Li,   $^7$Be,  $^{12}$C, $^{13}$C, $^{14}$N, $^{15}$N, $^{16}$O, $^{17}$O, $^{18}$O, $^{19}$F, $^{20}$Ne, $^{21}$Ne, $^{22}$Ne, $^{23}$Na,  $^{24}$Mg, $^{25}$Mg, $^{26}$Mg, $^{26}$Al, $^{27}$Al, $^{28}$Si, 
$^{32}$S ,$^{36}$Ar,   $^{40}$Ca,    $^{44}$Ti,    
$^{48}$Cr,    $^{52}$Fe,    
$^{56}$Ni, and $^{60}$Zn, \citep[see][]{2018MNRAS.476..496F,2021MNRAS.501.4514C}.

The major upgrade in the \textsc{parsec}~V2.0 is the inclusion of rotational effects with angular momentum transport and chemical mixing. Both processes are treated with purely diffusive schemes. 
The nuclear reactions and chemical variation equations are solved by the implicit method.
The abundance variation includes nuclear reactions, turbulent motions, rotational mixing, and molecular diffusion \citep{2019MNRAS.485.4641C}.
The extra mixing produced by thermohaline is included in the code by adding the corresponding thermohaline mixing diffusion coefficient (Eq. \ref{Dth})
to the turbulent one, which is included in the equation of chemical turbulent transport.

\textsc{parsec} accounts for mixing by convective overshooting beyond the Schwarzschild border, both in the core convective region and at the bottom of the convective envelope. While the extension of the core overshooting region is determined by means of a ballistic approximation scheme \citep{Bressan1981A&A}, the length of the overshooting region below the envelope convective region is parameterized as $L_{EOV}=\Lambda_{\mathrm{e}}~H_P$, with $H_P$ being the pressure scale height and 
$\Lambda_{\mathrm{e}}$ the envelope overshooting parameter \citep{1991A&A...244...95A}. 
Usually, the value adopted for $\Lambda_\mathrm{e}$ depends on the initial mass, being  $\Lambda_\mathrm{e}=0.5$ in low-mass stars and $\Lambda_\mathrm{e}=0.7$ in intermediate-mass stars with a linear transition between these two limits \citep{2012MNRAS.427..127B,2022A&A...665A.126N}. 
However, in this work, we let $\Lambda_\mathrm{e}$ as a free parameter that will be calibrated by comparing the models with observations (Sect.~\ref{EOV_calib}). In \textsc{parsec} standard model, $\Lambda_\mathrm{e}$, once it is indicated, is uniformly applied throughout the entire evolution of the stars. However, in Sect.~\ref{plateau}, an attempt to use this parameter differently between the early- and post-MS phases will be discussed in detail.

In the following sections, we will compute a series of preliminary models of a typical low metallicity, Z=0.0002, low mass star, M$=0.78\Msun$, with different adopting values of $\alpha_\mathrm{th}=50, 100, 150$ to analyse the efficiency of thermohaline mixing, and its impact on the stellar evolution. We will also adopt two different values of $\Lambda_{\mathrm{e}}=0.5$ and $\Lambda_{\mathrm{e}}=0.6$ to study the combined effects of envelope overshooting and thermohaline mixing.

\subsection{The thermohaline instability region}\label{TH_boders}
\begin{figure}[t]
    \includegraphics[width=\columnwidth]{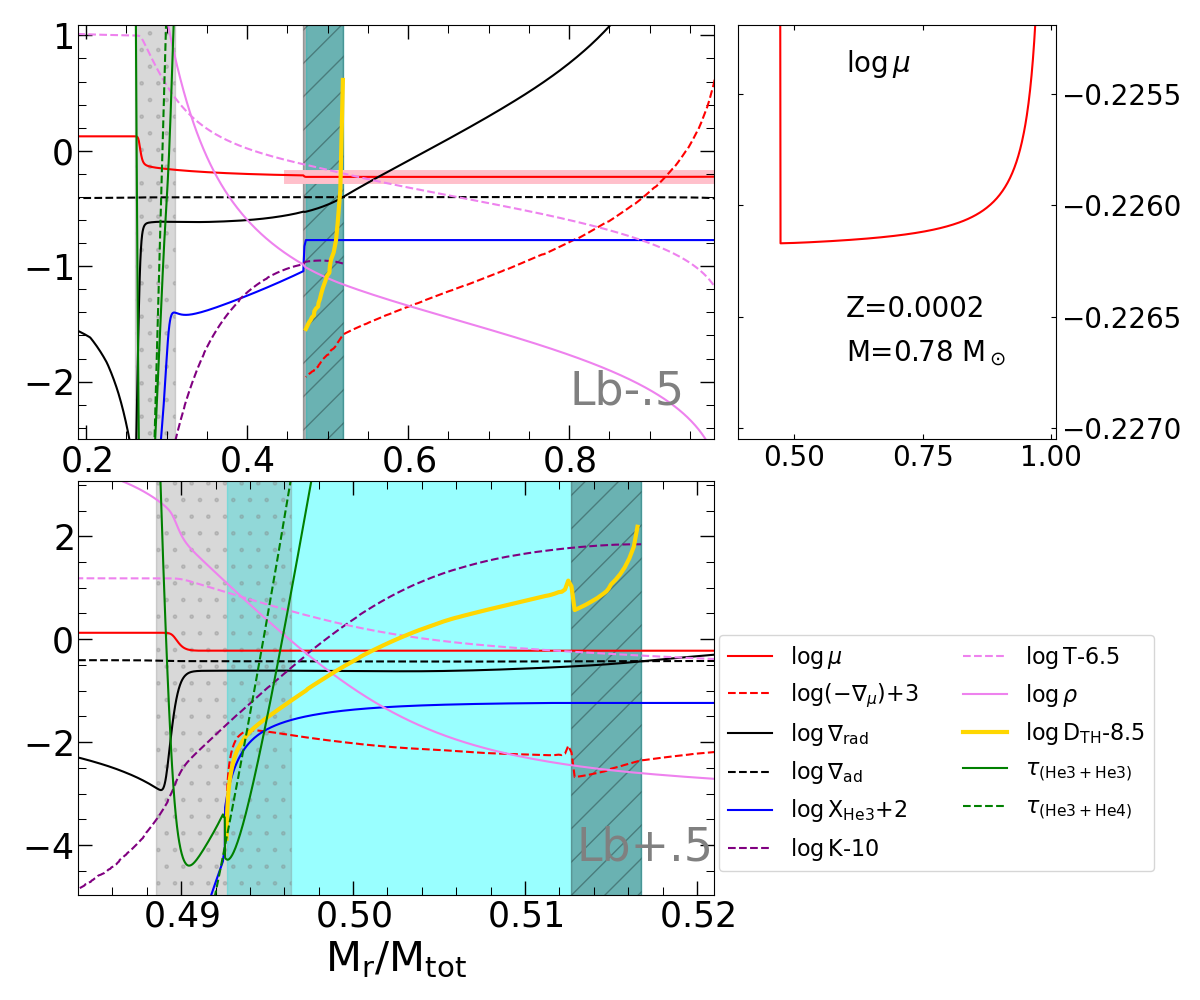}
    \caption{Internal structure of the model with   Z$=0.0002$, M$=0.78\Msun$, $\Lambda_\mathrm{e}=0.6$ and $\alpha_\mathrm{th}=50$. \textit{Top-left} and \textit{bottom-left} panels are the illustrations at the point below and above the RGB-bump luminosity (Lb) $0.5$dex. \textit{Top-right} panel is the zoom-in to the structure of $\log\mu$ at the outer layers that is marked by the pink-band on the top-left panel. The cyan shaded-area indicates the thermohaline instability. The grey-dashed-area indicates the envelope overshooting region. The grey-dotted-area indicates the H-burning shell. The green lines are the decay timescale of different reactions.}
    \label{mu_var}
\end{figure}

Figure \ref{mu_var} depicts the run of the internal main quantities related to the thermohaline instability region at two evolutionary points along the RGB phase, i.e. at a luminosity $0.5$ dex dimmer and brighter than that of the RGB bump, respectively. 

In the stellar model located below the RGB bump (\textit{top-left} panel), the mean molecular weight gradient, plotted in the top-left panel as $-\nabla_\mu$ with a red-dashed line, is negative from the outer atmosphere layers till deep inside the star, at $M_\mathrm{r}/M_\mathrm{tot} \sim 0.47$, where the discontinuity in chemical distribution is located. Note that the chemical discontinuity, marked by the $^3\mathrm{He}$~profile, is located below the Schwarzschild border indicated by the run of the radiative and adiabatic gradients, because of envelope overshooting which is illustrated by the grey-dashed-area. In these chemically homogeneous layers, $\nabla_\mu$ is negative because of the increasing ionisation degree. 
Below the chemical discontinuity, at about $M_\mathrm{r}/M_\mathrm{tot} \sim 0.47$ (see top-right panel), $\mu$ jumps to higher values, and $\nabla_\mu$ becomes positive. In this model, the layers where the thermohaline diffusion coefficient, plotted by the yellow solid line, is larger than 0 are located between the chemical discontinuity and the  Schwarzschild border, above which the region is convectively unstable. The borders of the thermohaline region are indicated by the cyan-shaded area in Fig.~\ref{mu_var}. However, it is important to note that if this region corresponds to the overshooting one, thermohaline mixing hence has no effects. As can be seen, in this case, the two regions are superimposed. 

In the model after the RGB bump (\textit{bottom-left} panel), the H-burning shell has already erased the chemical discontinuity left by the first dredge-up. The region of efficient nuclear shell burning (grey-dotted area), marked by the step decrease of the $^3\mathrm{He}$ abundance, is much below the bottom envelope Schwarzschild border and cannot be reached even from the overshooting mixing. However, in this region, the molecular weight is 
also sensitive to the $^3\mathrm{He}$ destruction by the \He(\He,2p)$^4\mathrm{He}$ and \He($^4\mathrm{He}$,$\gamma$)$^7\mathrm{Be}$ reactions. 
Since, as shown in the figure, the reaction \He(\He, 2p)$^4\mathrm{He}$ (solid-green line) is faster than the \He($^4\mathrm{He}$,$\gamma$)$^7\mathrm{Be}$ one,  
an extended stratification with negative $\nabla_\mu$ is produced 
between the $^3\mathrm{He}$ burning shell and the bottom of the envelope convective region. Thus, because of the thermohaline instability and envelope convection, ashes of the H-shell nuclear reactions can reach the stellar photosphere.

\subsection{Effect of different efficiency values of $\alpha_\mathrm{th}$}
\begin{figure}[t]
    \centering
    \includegraphics[width=\columnwidth]{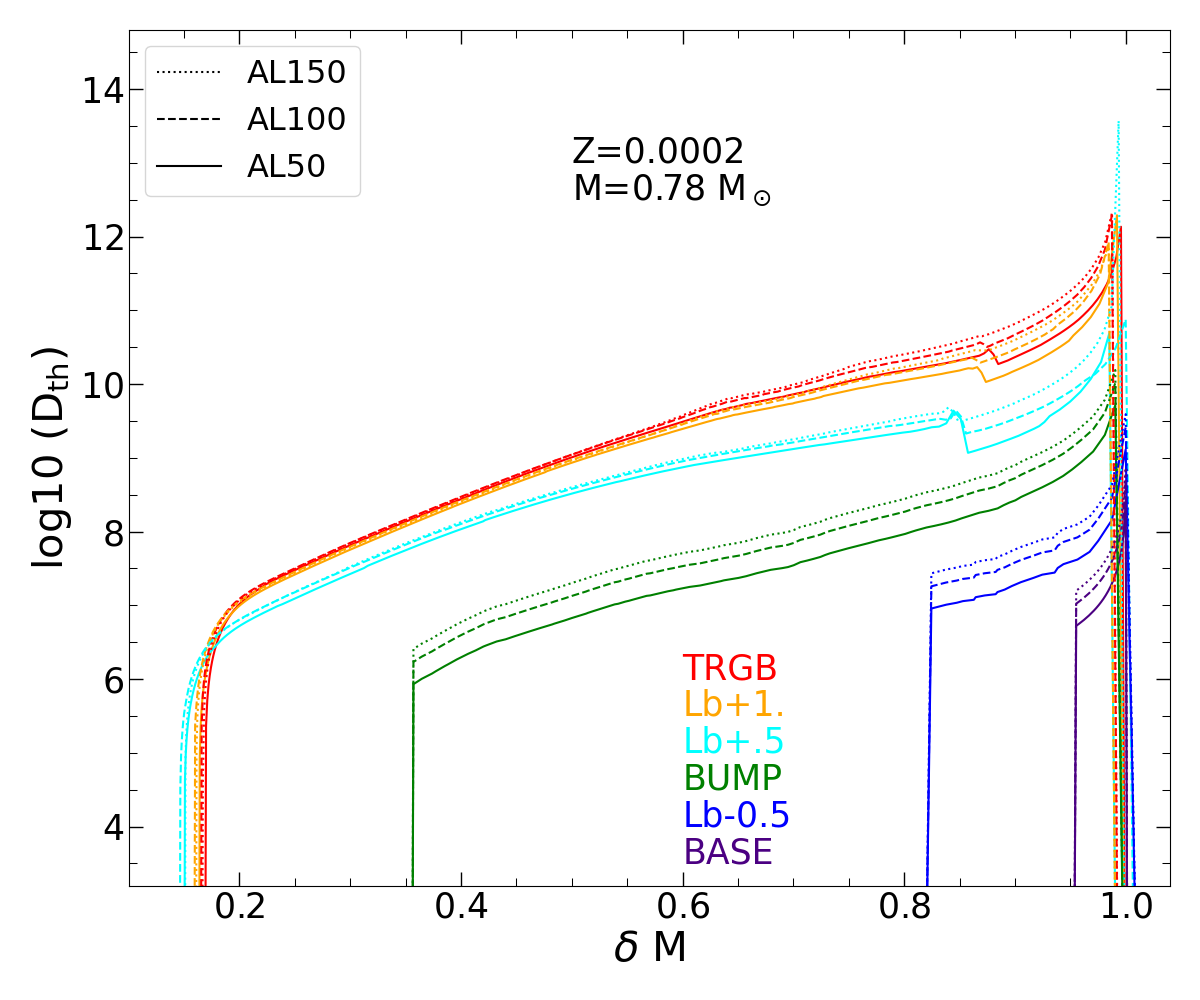}
    \caption{Variation of thermohaline diffusion coefficient within the radiative region, at five typical stages along the evolution of a selected star $0.78\Msun$. The x-axis is the relative mass in Eq. \ref{delM}. Models with different efficiency parameters, $\alpha_\mathrm{th}=50, 100, 150$, (referred as \protect\say{AL}) are displayed with different line styles, while colors indicate the five stages: at the base of the RGB (BASE), $0.5$ dex below the bump's luminosity (Lb-0.5), at the bump's luminosity (BUMP), $0.5$ dex above the bump (Lb+0.5), $1$ dex above the bump (Lb+1) and at the tip of the RGB (TRGB).}
    \label{Dth_ALs}
\end{figure}

We recall that in the current investigation, we have investigated the effects of the thermohaline mixing process only in the H-rich envelopes of RGB models.
To see how the efficiency of the thermohaline mixing process varies along the RGB, we compare in Fig.~\ref{Dth_ALs} the internal variation of $D_\mathrm{th}$ at six typical stages along the evolution of the calculated models. 
The x-axis represents the mass fraction above the H-exhausted core relative to the mass of the inter-shell between the core and the convective envelope
\begin{align}\label{delM}
    \delta M = \frac{M_r - M_\mathrm{c}}{M_\mathrm{env} - M_\mathrm{c}},
\end{align}
with $M_r$ the mass coordinate and  $M_\mathrm{env}$, $M_\mathrm{c}$ are the values at the base of the convective envelope, and where the hydrogen $X_c<10^{-7}$, respectively. 
Namely, $\delta M=0$ marks the border of the inner core while $\delta M=1$ marks the base of the envelope.

One can see that, as the star ascends the RGB, the thermohaline mixing process becomes more efficient, with $D_\mathrm{th}$ increasing by order of magnitudes, with   
$\alpha_\mathrm{th}$ modulating its efficiency. 
The growth of $D_\mathrm{th}$ is due to the rapid increase of thermal diffusivity $K$ due to the decrease of the mass density during the evolution of red giant stars. For example, between the two locations of $0.5$ dex below and above the bump, thermohaline diffusion coefficient changes of $1.56$ dex at the bottom of the envelope, and the thermal diffusivity $K$ increases $2.8$ dex (as shown in Fig.~\ref{mu_var}). This can be understood by the inverse relation of thermal diffusivity with density. In other words, the expansion during the giant phase leads to a significant decrease in the density and hence to a rise of the thermal diffusivity, and so of $D_\mathrm{th}$. 
Another interesting point is that the instability region grows in size (in terms of relative mass), from the base of the envelope to the H-burning shell region, along the evolution of the stars. Finally, the strength of thermohaline depends directly on its efficiency parameter, namely, the larger  $\alpha_\mathrm{th}$, the higher $D_\mathrm{th}$.

\begin{table}[!tbp]
\caption{Location and size of the RGB bump, together with the luminosity and age at the tip of the RGB of a given model with $0.78\Msun$, $Z=0.0002$ and for four values of $\alpha_\mathrm{th}=0-150$, and two sets of models with $\Lambda_\mathrm{e}=0.5$ and $0.6$.}
\label{tab1}
\centering
\begin{tabular}{l c c c c}
\hline\hline
 & $\log L_\mathrm{avg}$ & $\Delta \log L_\mathrm{bump}$ & $\log L_\mathrm{tip}$ & $t_\mathrm{tip}$ (Gyr) \\
\hline
\multicolumn{5}{c}{$\Lambda_\mathrm{e}=0.5$} \\
\hline
$\alpha_\mathrm{th}=0$ & 2.04906 & 0.00622 & 3.34714 & 13.43002 \\
$\alpha_\mathrm{th}=50$ & 2.05257 & 0.00776 & 3.33226 & 13.42805 \\
$\alpha_\mathrm{th}=100$ & 2.05470 & 0.00948 & 3.33211 & 13.42798 \\
$\alpha_\mathrm{th}=150$ & 2.05591 & 0.01072 & 3.33128 & 13.42792 \\
\hline
\multicolumn{5}{c}{$\Lambda_\mathrm{e}=0.6$} \\
\hline
$\alpha_\mathrm{th}=0$ & 2.02502 & 0.00673 & 3.34710 & 13.43035 \\
$\alpha_\mathrm{th}=50$ & 2.02877 & 0.00796 & 3.33321 & 13.42829 \\
$\alpha_\mathrm{th}=100$ & 2.03070 & 0.00937 & 3.33188 & 13.42818 \\
$\alpha_\mathrm{th}=150$ & 2.03214 & 0.01114 & 3.33109 & 13.42811 \\
\hline
\end{tabular}
\end{table}
As a consequence, the additional mixing directly changes the distribution of chemical elements and, eventually, the further evolution of the stars during the RGB.

Concerning the differences in the HR diagram, we compare in Table. \ref{tab1} the average luminosity of the bump between the maximum and minimum values ($\log L_\mathrm{avg}/L_\odot$), the size of the bump in terms of $\log L/L_\odot$ (meaning, the difference between the minimum and maximum luminosity of the bump), and the luminosity and age at the tip of the RGB phase. A larger value of $\alpha_\mathrm{th}$ leads to a slightly more luminous RGB bump and to a slightly larger difference between the minimum and maximum luminosity of the bump, which, in turn, takes to a slightly lower RGB tip luminosity and a younger RGB tip age. However, it should be emphasised that the differences here are rather modest. As expected, changing $\Lambda_\mathrm{e}$ has a more significant impact, especially on the location of the bump.

\subsection{Chemical abundances}\label{vari_3elements}

\begin{figure*}
    \includegraphics[width=\textwidth]{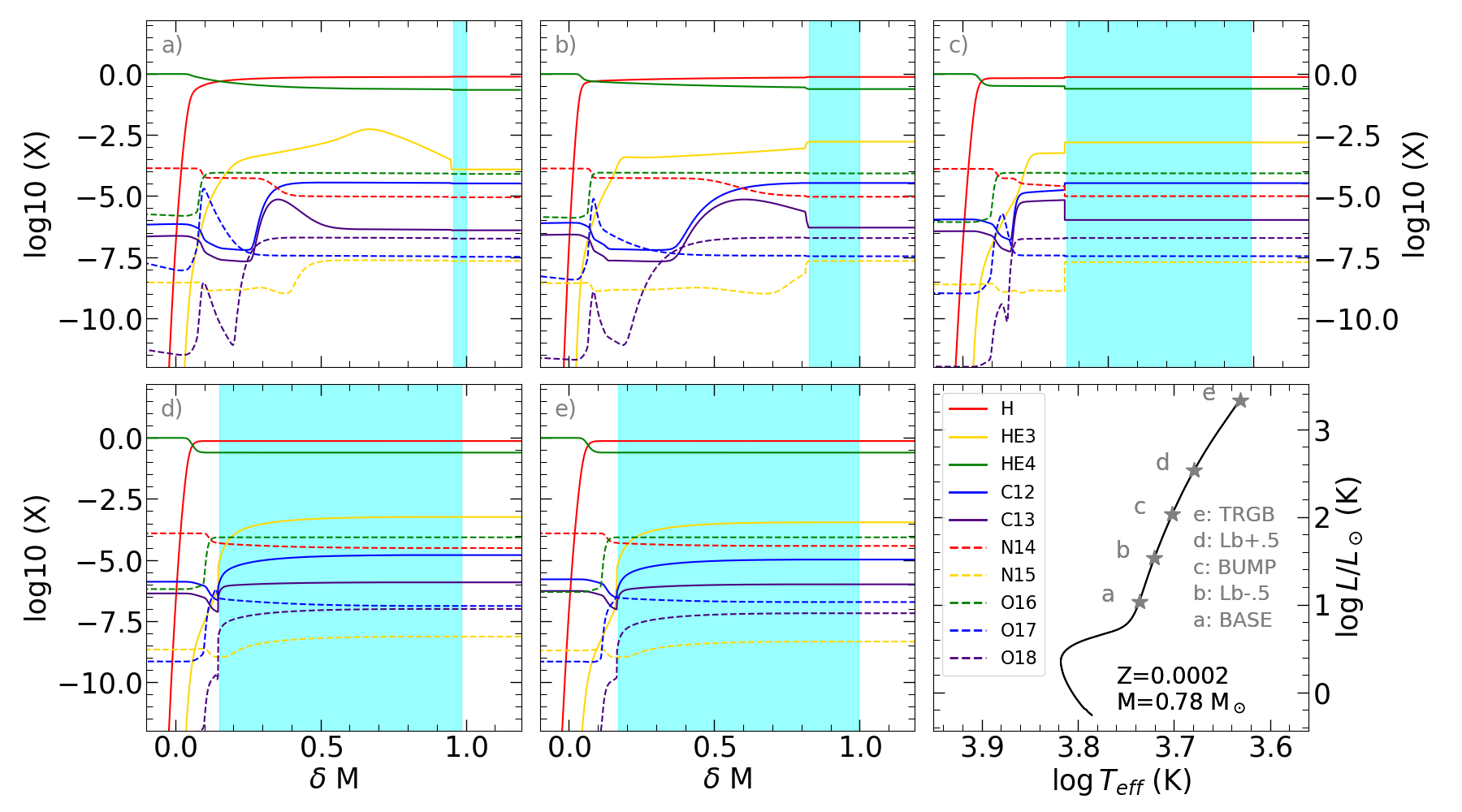}
    \caption[Variation of chemical elements in five stages along the RGB phase.]{Variation of chemical elements from $^1\mathrm{H}$ to $^{18}\mathrm{O}$ in five stages along the RGB phase of a selected star with initial mass $0.78\Msun$ and metallicity $Z=0.0002$. The HRD of the corresponding star begins from the zero-age-main-sequence, with the five marked stages. Each element is indicated by the coloured lines. The cyan-shaded area displays the thermohaline instability region at each evolutionary stage (bottom-right panel). All first five panels share the same scale on the x- and y-axis.}
    \label{chemi_variation}
\end{figure*}

Since thermohaline mixing can convey elements from the inner H-shell to the base of the outer convective envelope, its largest effect is expected to appear in their abundance evolution, particularly at the stellar surface.
In Fig.~\ref{chemi_variation}, we show the structure of 10 chemical elements from $^1$H to $^{18}$O in the radiative zone between the central core and the envelope of a star $0.78\Msun$, $Z=0.0002$, at the same five stages of Fig.~\ref{Dth_ALs}.

At the base of the RGB phase, $\He$ accumulates as a result of the pp-chain nuclear reaction from the main sequence and reaches its peak in the shells above the central core. It is brought up to the convective envelope at first and then transported down to the region of the H-burning shell due to the 1DU later on.

In the deeper layers, the variation of $^{12}$C and $^{13}$C result from the competition between $^{12}$C($^1$H,$\gamma$)$^{13}$N(e$^+$+$\nu_\mathrm{e}$)$^{13}$C and
$^{13}$C($^1$H,$\gamma$)$^{14}$N($^1$H,$\gamma$)$^{15}$O(e$^+$+$\nu_\mathrm{e}$)$^{15}$N($^1$H,$^4$He)$^{12}$C
via the CNO-I cycle. In deeper layers, the depletion of $^{17}$O and $^{18}$O results from the ON-cycle that enriches $^{16}$O and creates its plateau. The variation of $^{14}$N and $^{15}$N is also a consequence of the competition between those nuclear processes.

Further ascending to the RGB bump at which the penetration of the convective envelope has its maximum extension, and $\He$ is already been engulfed down to the H-burning shell regions. Up to this point, the inner temperature is high enough to ignite the $^3$He reactions and enlarges the region of thermohaline instability. 
During the further evolution up to the tip of the RGB, the efficient thermohaline region increases in size, mixing the inner layers above the H-burning shell with the external envelope. As a consequence, thermohaline mixing enriches the abundance of $^{14}$N while $^{12}$C and $^{13}$C and $^{15}$N are depleted.
Overall throughout the whole RGB phase, thermohaline mixing results in a depletion of the surface abundance of $\He$, $^7$Li,$^{12}$C, $^{13}$C, $^{15}$N, $^{18}$O, $^{19}$F, $^{22}$Ne, and an increase of $^{14}$N, $^{17}$O, $^{23}$Na, while the surface abundances of other heavier elements remain unaffected because they are not involved in nuclear reactions at this early stage of evolution.

\begin{figure*}
    \includegraphics[width=\textwidth]{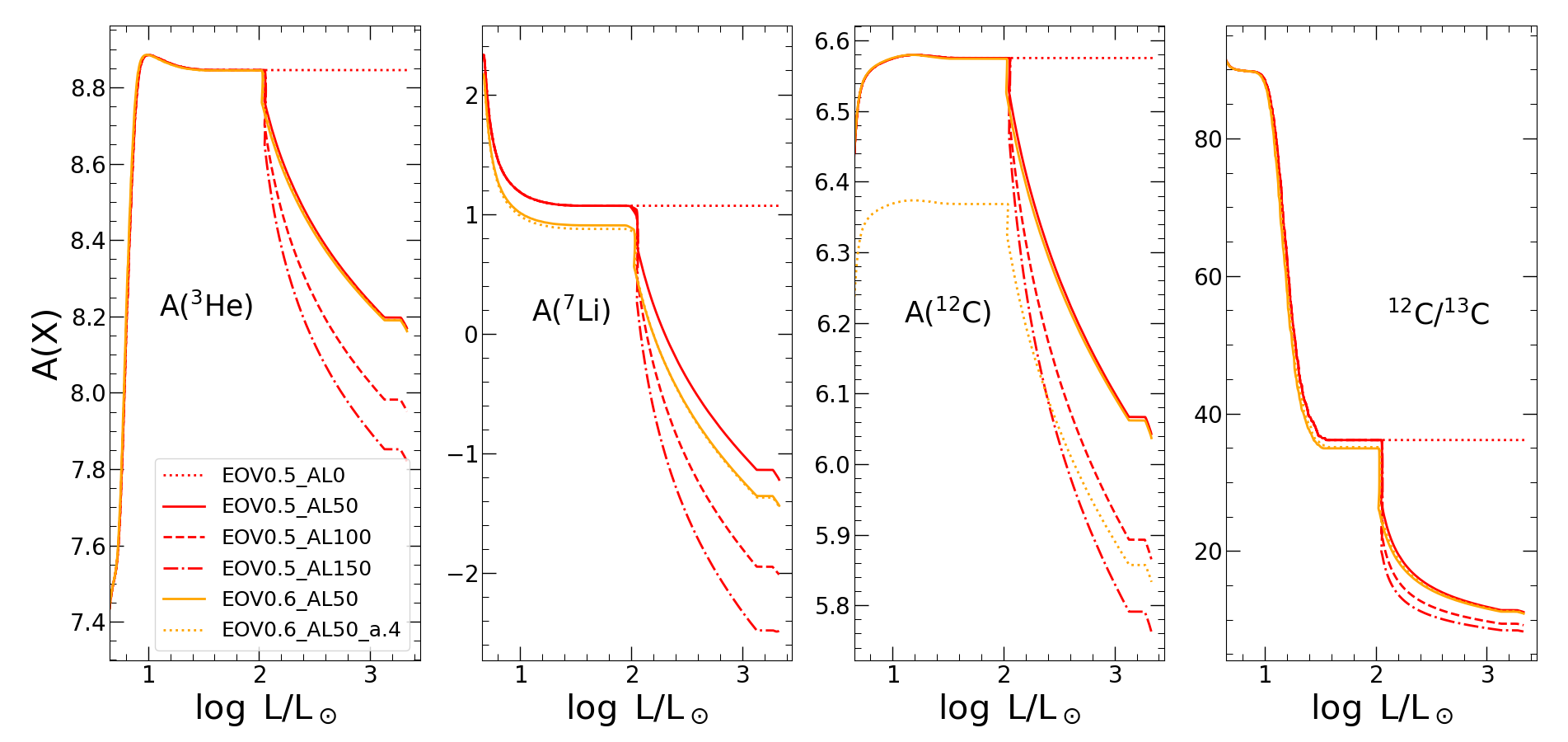}
    \caption{Comparison between different models on the variation surface abundances of $^3$He, $^7$Li, $^{12}$C and C-isotopes ratio. The models with different values of $\Lambda_\mathrm{e}=0.5,0.6$, and $\alpha_\mathrm{th}=0, 50, 100, 150$ by using solar-scaled and [$\alpha$/Fe]=0.4 mixtures are shown according to the indicated labels. For simplicity, only the evolution during the giant phases is shown here.}
    \label{chemi_single_mass_vari}
\end{figure*}

Of particular interest is the surface abundance evolution of $^3\mathrm{He}$, $^7\mathrm{Li}$, $^{12}\mathrm{C}$ and $^{13}\mathrm{C}$ during the RGB phase. The first element drives the thermohaline process, while the abundance of the other elements is commonly used to probe the internal mixing of low-mass giant stars by comparing their observed values.

Figure~\ref{chemi_single_mass_vari}
shows the evolution of the surface abundances of $^3$He, $^7$Li, $^{12}$C\footnote{$\mathrm{A(X)}=\log(\mathrm{n}_\mathrm{X}/\mathrm{n}_\mathrm{H})+12$} and C-isotopic ratio\footnote{$\mathrm{^{12}C/^{13}C}=\mathrm{n_{^{12}C}/n_{^{13}C}}$}, as a function of the stellar luminosity for the star with $M=0.78\Msun$, $Z=0.0002$ and different assumptions of the envelope overshooting and thermohaline parameters, as indicated in the caption.

The first hump of the $^3\mathrm{He}$ abundance at $L\sim 10\Lsun$ is the result of 1DU. In correspondence to this hump, we note a strong decrease of $^7$Li, an increase of $^{12}$C, and a plateau in the $^{12}$C/$^{13}$C abundance ratio. 
The following variation up to $L\sim 20\Lsun$ is produced by the further penetration of envelope convection up to its maximum depth. Since then, surface abundances remain unchanged while, in the meantime, the H-burning shell advances until it reaches the discontinuity left by the maximum penetration of envelope convection. This happens at the RGB bump at $L\sim 100\Lsun$ for the adopted models. Worth to note is the barely visible variation of the RGB bump luminosity and the more conspicuous effect on A($^7\mathrm{Li}$) and $^{12}$C/$^{13}$C ratio, produced by the different assumptions of envelope overshooting efficiency. These points will be discussed in more detail in Sect. \ref{alpha_th_calib}.

As the H-shell moves forward to the discontinuity, which remains very near to the base of the convective envelope but without thermohaline circulation (model EOV0.5\_AL0, red dotted line), there will be no further 
mixing between these two regions.
Instead, in the presence of thermohaline, the mixing is allowed with an efficiency that strongly depends on the thermohaline efficiency parameter \citep{2007A&A...467L..15C,2010A&A...521A...9C}.
Finally, we can see how the variation of $^{12}$C is due to mixing and comparing with its variation due to the $\alpha$-enhancement.

From the above discussion, we conclude that the envelope overshooting efficiency establishes the location of the H discontinuity in the interior structure, and thus the luminosity of the RGB bump. It also affects the surface abundance of Li at the RGB. On the other hand, the efficiency of thermohaline mixing drives the subsequent depression of A($^7\mathrm{Li}$), A($^{12}\mathrm{C}$) and of the $^{12}\mathrm{C}/^{13}\mathrm{C}$ ratio \citep[see also][and references therein]{2017MNRAS.469.4600H, 2023A&A...670A..73A}.  
In the next sections, we will use these two different effects to disentangle the two processes and to calibrate the thermohaline and envelope overshooting efficiency parameters.

\section{Distance independent calibration of $\alpha_\mathrm{th}$ and $\Lambda_\mathrm{e}$ from Li-abundance in NGC 6397}\label{alpha_th_calib}

In order to constrain the thermohaline efficiency parameter $\alpha_\mathrm{th}$ in Eq.~\ref{Dth}, we compare our model predictions with the $^7$Li abundance of 349 stars of the metal-poor GC NGC 6397 from \citet{2009A&A...503..545L}, and the $^{12}$C abundances from \citet{1990ApJ...359..307B}. Furthermore, the relative distance between the Li-peak on the subgiant and the location of the RGB bump implies a robust calibrator for envelope overshooting. These will be discussed in detail below.

\subsection{Li-abundances and thermohaline mixing}

NGC 6397 is one of the most metal-poor GCs. Determination of its metallicity has been pursued in many works, for instance, \citet{2003A&A...408..529G} estimates the metallicity $\feh=-2.03\pm 0.05$, \citet{2009A&A...503..545L} determines $\feh=-2.10$ while \citet{2007ApJ...671..402K} find $\feh=-2.28\pm 0.04$ for turn-off stars and $\feh=-2.12\pm 0.03$ for RGB stars the difference being attributed to atomic diffusion. 
The distance modulus of NGC 6397 has also been studied in many individual works, from CMD fitting to kinematic simulations and parallax analysis. For instance, \citet{1998AJ....116.2929R,2003A&A...408..529G,2007ApJ...671..380H,2020JCAP...12..002V,2023MNRAS.526.5628G} report an estimated distance of about $\sim 2.46-2.67$ kpc by using CMD fitting. Also with the CMD fitting method, \citet{2018ApJ...864..147C} derive a distance of $12.02\pm 0.03$ mag ($\sim 2.53$ kpc). The N-body model-fitting method performed in \citet{2019MNRAS.482.5138B} for 154 Galactic globular clusters gives a distance of $2.45\pm 0.04$ kpc for NGC 6397. A trigonometric parallax analysis on the cluster by \citet{2018ApJ...856L...6B} gives a distance of $2.39\pm 0.17$ kpc taking into account contributions from both statistical and systematic errors, or \citet{2021MNRAS.505.5957B} who give a mean distance of $2.482\pm 0.019$ kpc. 
The age of this cluster is estimated to lie between 12.5-13.5 Gyr within the estimated uncertainties \citep[see][]{2003A&A...408..529G,2010ApJ...708..698D,2013ApJ...775..134V,2018ApJ...864..147C,2020MNRAS.498.5745T,2021NewA...8801607A}. The turn-off mass is about $0.78\Msun$ \citep[see][]{2007ApJ...671..402K}.

\begin{figure}[t]
    \centering
    \includegraphics[width=\columnwidth]{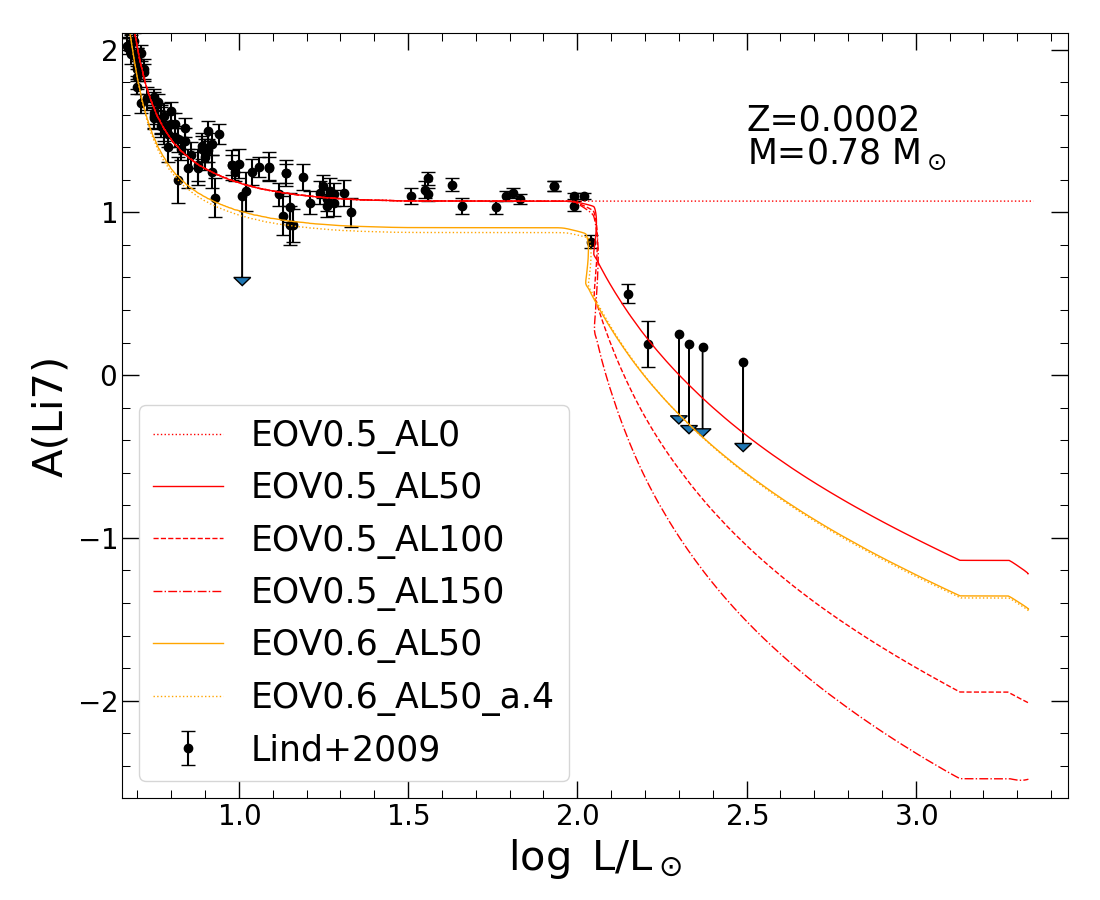}
    \caption{Variation of $\Li$ abundance versus luminosity of giant stars. The computed model with $\Lambda_\mathrm{e}=0.5$ and solar-scaled mixture is shown in red, with four values of $\alpha_\mathrm{th}=0, 50, 100, 150$ showed by different line styles. The yellow lines show models with $\Lambda_\mathrm{e}=0.6$, $\alpha_\mathrm{th}=50$ using solar-scaled mixture (solid) and [$\alpha$/Fe]=0.4 mixtures  (dotted).}
    \label{Ali_M0.78_Z0.0002}
\end{figure}

The lithium abundance of in total 349 stars spanning from the MS turn-off point to the RGB with 1D+NLTE corrections is adopted from \citet{2009A&A...503..545L} and is shown in Fig.~\ref{Ali_M0.78_Z0.0002}. Among 90 RGB stars, there are seven stars located at the luminosity above the RGB bump, which requires an extra-mixing to be reproduced \citep[see][]{2007ApJ...671..402K,2017MNRAS.469.4600H}. 
In Fig.~\ref{Ali_M0.78_Z0.0002}, we compare with the data the variation of A$(\Li)$ along the evolution of stellar models of $M=0.78\Msun$ and $Z=0.0002$, which is appropriate for the early post-MS evolution of NGC 6397.
While it is well known that the 1DU event is responsible for the first falling branch in the low-luminosity domain around the RGB base in Fig.~\ref{Ali_M0.78_Z0.0002}, in standard models (with no extra-mixing and $\alpha_\mathrm{th}=0$, model EOV0.5\_AL0) the abundance of $\Li$ remains constant after the bump since there is no further mixing. However, in the presence of thermohaline mixing, $^7$Li diffuses downward from the envelope leading to the second falling branch above the RGB bump. Here, we would like to stress that it is difficult to draw firm conclusions from the four most luminous stars in \citeauthor{2009A&A...503..545L} sample because they only have upper limits on the $\Li$ abundance. Therefore, we chose the three stars that are nearest to the bump as our best-fit constraint. Indeed, the lithium abundance of these three stars is very well predicted by our model with $\alpha_\mathrm{th}=50$.

\begin{figure}[t]
    \centering
    \includegraphics[width=\columnwidth]{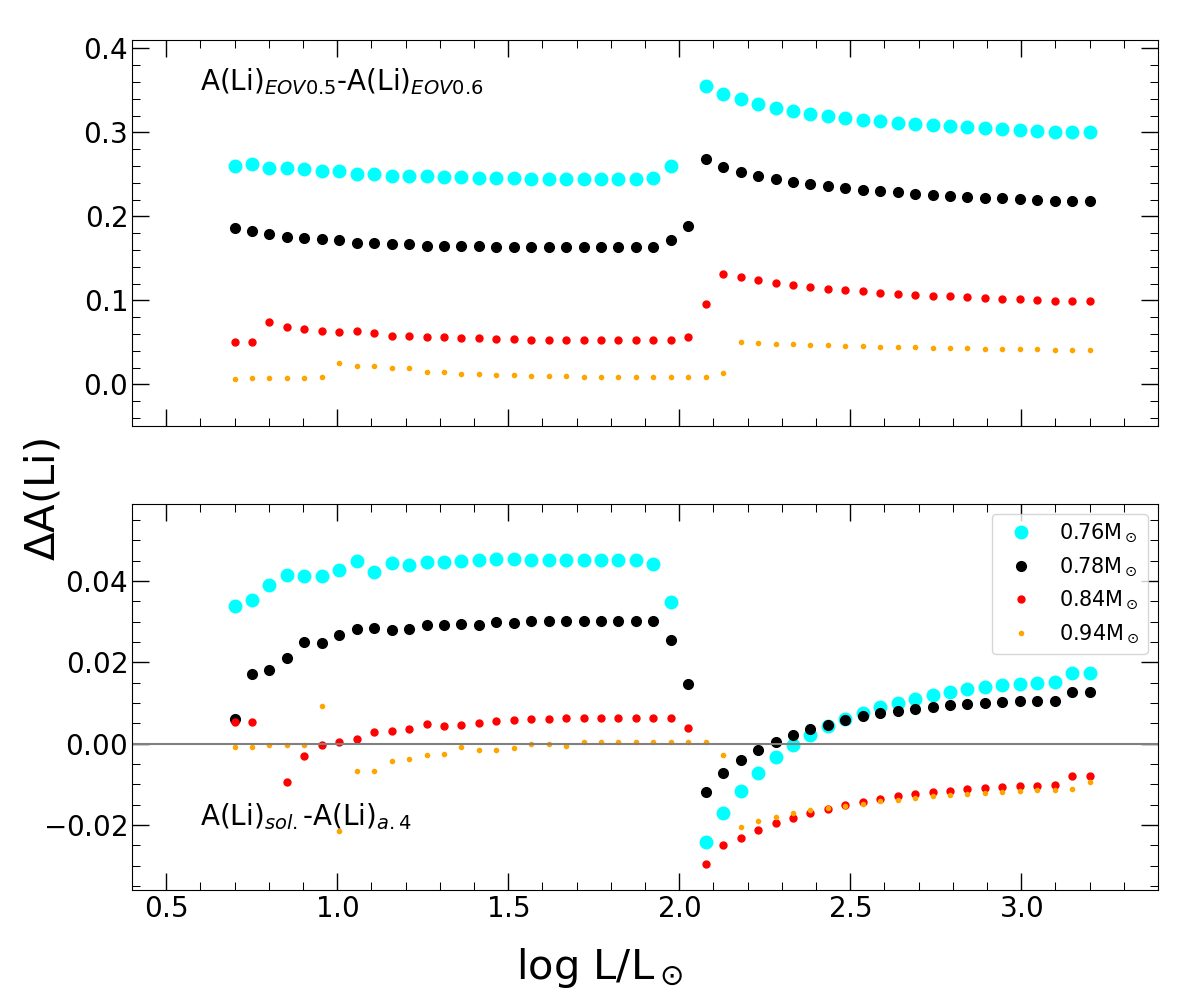}
    \caption{A($\Li$) differences between models in Fig.~\ref{Ali_M0.78_Z0.0002}. \textit{Upper panel}: between model with $\Lambda_\mathrm{e}=0.5$ and $\Lambda_\mathrm{e}=0.6$, by using the solar-scaled mixtures. \textit{Bottom panel}: between model using solar-scaled and $\alpha$-enhanced mixtures, with the same value $\Lambda_\mathrm{e}=0.6$.}
    \label{D_Ali}
\end{figure}

Besides that, in Fig.~\ref{Ali_M0.78_Z0.0002}, we show the models that are calculated with envelope overshooting efficiency parameter $\Lambda_\mathrm{e}=0.5$ which is the value used in \textsc{parsec} v2.0 models, and a higher value $\Lambda_\mathrm{e}=0.6$. As can be seen, $\mathrm{A}(^7\mathrm{Li})$ changes significantly by changing the efficiency of envelope overshooting. To the contrary, the variation of $\Li$ abundance is very similar between models computed with solar-scaled and $\alpha$-enhanced mixtures. To better understand this point, 
Fig.~\ref{D_Ali} shows the differences in A($\Li$) between models with $\Lambda_\mathrm{e}=0.5$ and models with $\Lambda_\mathrm{e}=0.6$ (top panel), and between models with solar-scaled and $\alpha$-enhanced mixtures (bottom panel) for a few models with given initial mass. In the case of the $0.78\Msun$ model, a minimum of $\sim 0.16$\,dex difference is found when varying $\Lambda_\mathrm{e}$ from $0.5$ to $0.6$ at the early phase below the bump. The discrepancy tends to be larger after the bump with a maximum difference of about $\sim 0.3$\,dex. The absolute value of the change also depends on the initial mass, as shown in the figure. Namely, the higher the mass, the smaller the difference.
On the other hand, the maximum difference between the two models of using solar-scaled and $[\alpha/\mathrm{Fe}]=0.4$-mixtures is $\sim 0.04$\,dex, in all models, as can be seen at the bottom panel.

For a more refined calibration, we compute several low-mass models, in the range of masses $M=0.64-0.94\,\Msun$, for two initial metallicity sets $Z=0.0001, 0.0002$, with the $[\alpha/\mathrm{Fe}]=0.4$ mixture. For each metallicity set, we compute models with both values of $\Lambda_\mathrm{e}=0.5$ and $0.6$ and a value of $\alpha_\mathrm{th}=50$. All models are set up with the initial lithium $\mathrm{A(^7Li)}=2.72$. Additionally, we also compute a set with $\Lambda_\mathrm{e}=0.6$ and $\alpha_\mathrm{th}=100$, and a set with $\Lambda_\mathrm{e}=0.05$ and $\alpha_\mathrm{th}=50$ for the sake of comparison.
Then, together with our newly computed models, we adopt the very-low-mass models ($\mathrm{M}\leq 0.60\,\Msun$) from the \textsc{parsec v1.2s} dataset\footnote{\url{http://stev.oapd.inaf.it/PARSEC/tracks_v12s.html}} to produce the corresponding isochrones. The isochrones are produced by the recent version of the \textsc{trilegal} code \citep[more details of the interpolation scheme can be found in][]{2005A&A...436..895G,2017ApJ...835...77M,1990A&AS...85..845B, 2008A&A...484..815B} and converted into the photometric magnitudes by using the YBC table\footnote{\url{http://stev.oapd.inaf.it/YBC/}} \citep[][]{2019A&A...632A.105C}.

\begin{figure}[t]
    \centering
    \includegraphics[width=\columnwidth]{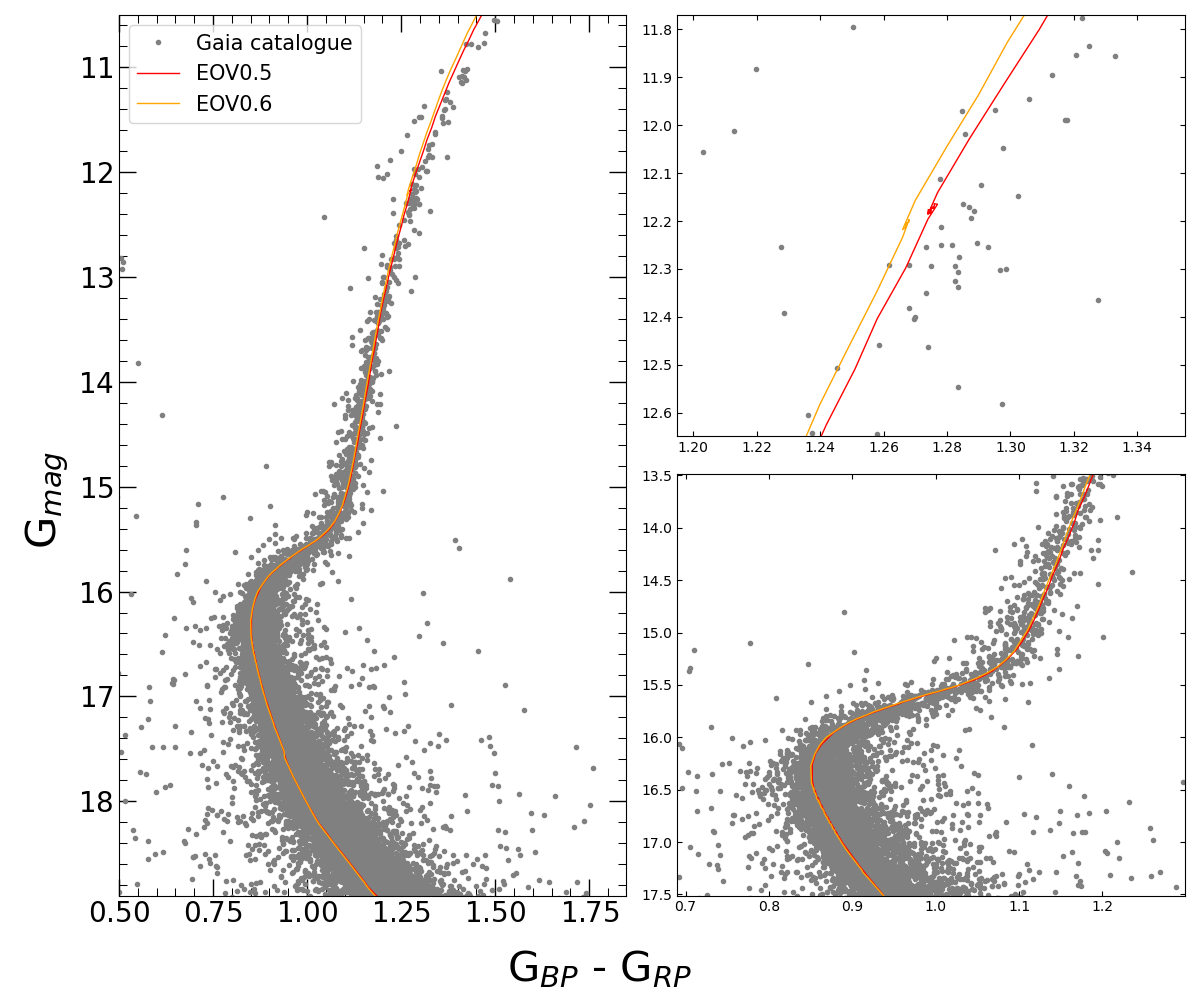}
    \caption{CMD of GC NGC 6397 in Gaia DR3 pass-bands. The data is taken from Gaia EDR3 catalogue \citet{2021MNRAS.505.5978V}. The isochrones are shown with the adopted parameters $(m-M)_0=12.05$ mag, $\mathrm{[Fe/H]}=-2.088$, $A_v=0.57$ mag and age $t=12.8$ Gyr. Two panels on the right-hand side are zoom into the bump and main-sequence areas.}
    \label{ngc6397_cmd}
\end{figure}
\begin{figure}[]
    \centering
    \includegraphics[width=\columnwidth]{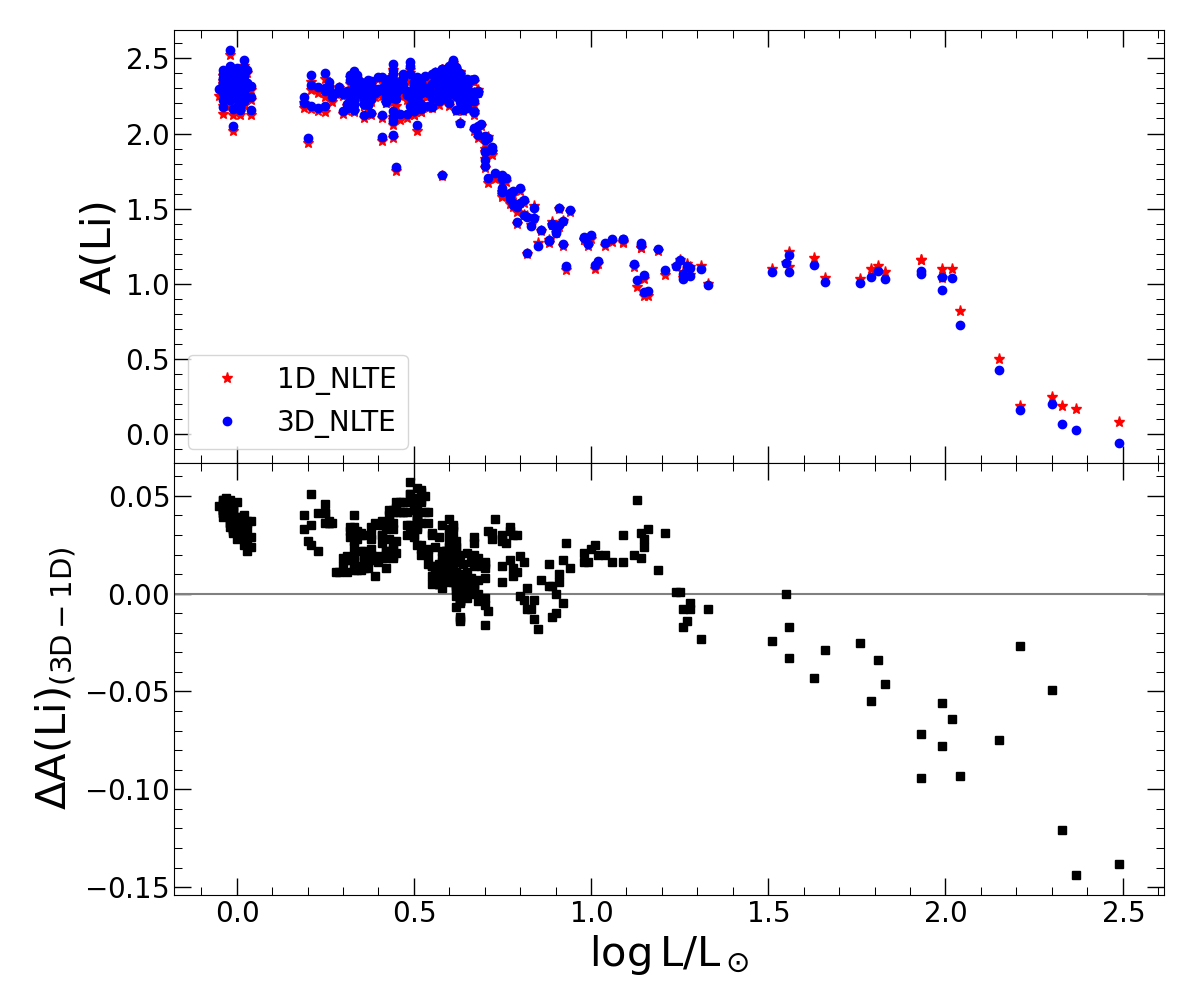}
    \caption{Comparison of $\mathrm{A}(^7\mathrm{Li})$ in GC NGC 6397 using 3D and 1D NLTE corrections. Top-panel: the derived abundances of all stars in the Lind et al.\ sample. Bottom-panel: the difference of $\mathrm{A}(^7\mathrm{Li})$ between the two methods.}
    \label{Ali_3D_1D}
\end{figure}

As the first analysis step, Fig.~\ref{ngc6397_cmd} shows the observed CMD of GC NGC 6397 in Gaia DR3 passbands, superimposed by our isochrones. Note that, for simplicity, the isochrones are limited up to the tip of the RGB phase. In Fig.~\ref{ngc6397_cmd}, we adopt the distance modulus $(m-M)_0=12.05$\,mag from \citet{2018ApJ...864..147C}, and initial metallicity $Z=0.0002$ ($Y=0.249$, corresponds to $[\mathrm{Fe/H}]=-2.088$ with $\mathrm{[\alpha/Fe]}=0.4$). Our best-fit isochrones imply an extinction of $A_V=0.57$\,mag and an age of $12.8$\,Gyr, which are in good agreement with literature \citep[][and others]{2018ApJ...864..147C,2023MNRAS.526.5628G}. Overall, our isochrones perform well, yielding a global fit from the low-MS part up to the red-giant branch of the cluster. Furthermore, the location of the bump selects the best-fitting model with $\Lambda_\mathrm{e}=0.6$ and $[\alpha/\mathrm{Fe}]=0.4$-mixture.

\begin{figure}[t]
    \centering
    \includegraphics[width=\columnwidth]{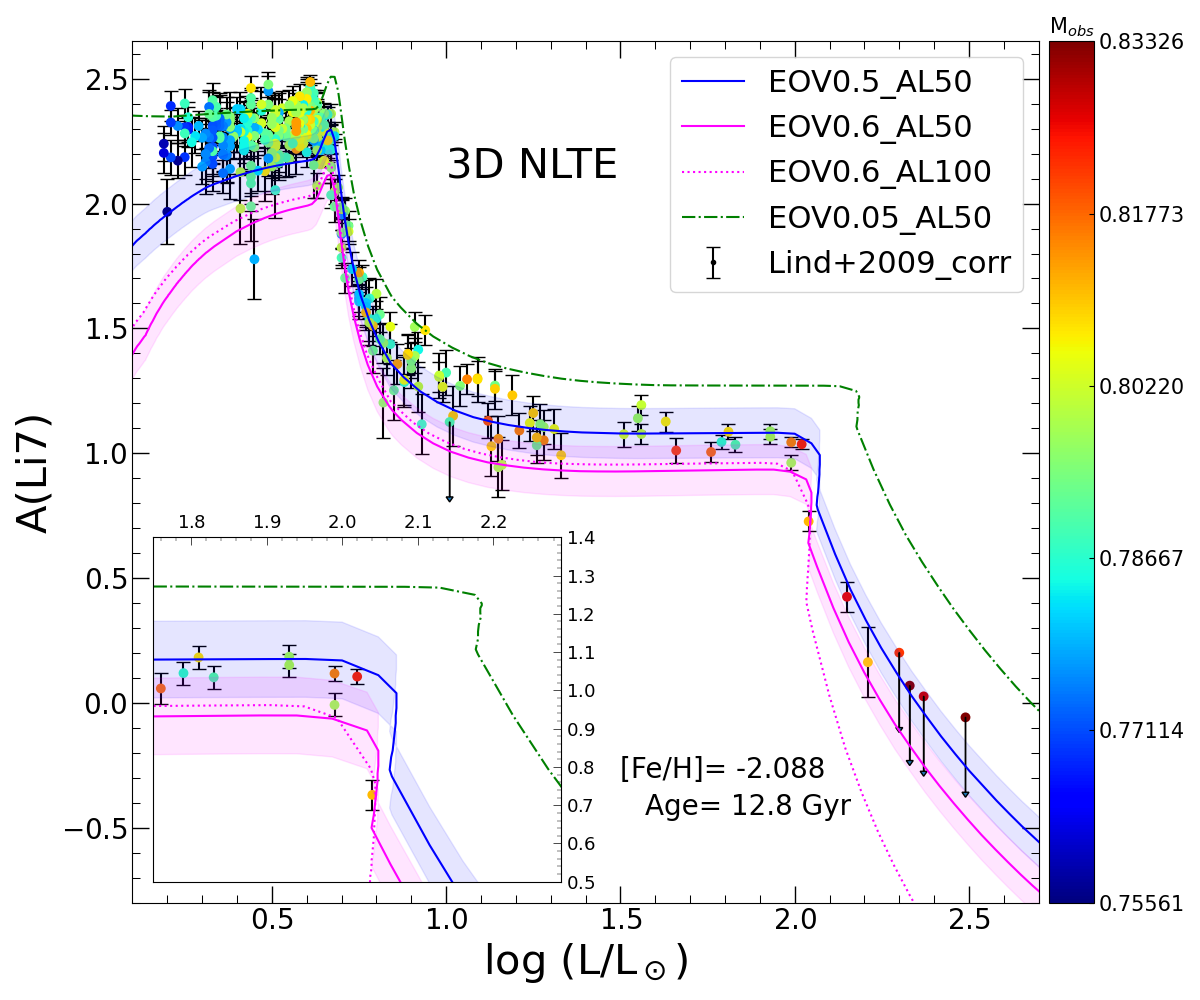}
    \caption{Isochrone fitting of the observed $\Li$-abundance data. The model giving the best CMD fit in Fig.~\ref{ngc6397_cmd} is compared to the 3D+NLTE-corrected data from \citet{2009A&A...503..545L}. A model with ($\Lambda_\mathrm{e}=0.6$, $\alpha_\mathrm{th}=100$) and one with ($\Lambda_\mathrm{e}=0.05$, $\alpha_\mathrm{th}=50$) are also plotted here for the sake of comparison. The insert panel zooms into the onset of thermohaline mixing. The colour bar indicates the observed masses.}
    \label{ngc6397_Ali_iso}
\end{figure}

After having constraints on the photometric properties from the CMD fit, we proceed to the comparison with the observed $\Li$-abundance trend. 
The reported 1D abundances from \citet{2009A&A...503..545L} are corrected for NLTE. Recently, \citet{2021MNRAS.500.2159W} have provided a grid of synthetic spectra and abundance corrections taking into account the combined effect from 3D+NLTE, which is provided in the \textsc{breidablik} package\footnote{\url{https://github.com/ellawang44/Breidablik}}. Taking advantage of this more recent update, we re-derived the abundance of all stars in the Lind et al.\ sample from their reported equivalent widths.
The comparison between data derived with 3D and 1D NLTE corrections is shown in Fig.~\ref{Ali_3D_1D} for all stars from the dwarf-MS to red-giants.
We find that for those stars below the RGB bump, there is an abundance shift of at most $\sim +0.05$\,dex between 3D and 1D NLTE corrections. For those stars located above the bump, the difference tends to increase up to a value of $\sim -0.15$\,dex. Most importantly, the corrections in the two domains have different signs, changing the overall trend. 

Figure~\ref{ngc6397_Ali_iso} shows the re-derived 3D+NLTE lithium abundances and the variation of $\mathrm{A}(^7\mathrm{Li})$ from our models versus luminosity. It should be noted that the models shown in Fig.~\ref{ngc6397_Ali_iso} are obtained from the best-fitting CMD above. We also include in the plot a model with $\alpha_\mathrm{th}=100$ for the sake of comparison with our constrained value $\alpha_\mathrm{th}=50$.
Investigating the comparison, one can see that the shape of lithium depletion after the RGB bump, indeed, is the one that verifies the constraint on the efficiency of thermohaline mixing. A reasonable uncertainty of $\pm 0.1$ dex due to the absolute value of $\mathrm{A(^7Li)}$ from the previous phases claims the best-fitted value is $\alpha_\mathrm{th}=50$. The model with $\alpha_\mathrm{th}=100$ enhances the efficiency of thermohaline mixing too much. It is also clear that high overshoot efficiencies engulf the surface lithium too much during the early MS; on the contrary, the observed data shows a well-known Spite-plateau. This point will be further discussed in the next subsection. 
Despite this, the location of thermohaline's onset shows its best-fit to the model with $\Lambda_\mathrm{e}=0.6$.

Finally, for comparison, our best-fitting thermohaline efficiency parameter is lower than the result of \citet{2017MNRAS.469.4600H} who gets $C_\mathrm{t}=150$, while our results indicate $\alpha_\mathrm{th}=50$ which corresponds to $C_\mathrm{t}=75$. However, we notice that their models use an envelope overshooting value of $0.14H_\mathrm{p}$ while in this work we calibrate and use a higher value, i.e., either $\Lambda_\mathrm{e}=0.5$ or $0.6$.

\subsection{Spite plateau}\label{plateau}
The predicted amount of lithium formed in the early universe is up to 3 times larger than the observed abundance in halo field and globular-cluster stars \citep[e.g.,][]{2005A&A...442..961C, 2002A&A...395..515B, 2007ApJ...671..402K}. For instance, \citet{1982A&A...115..357S,1982Natur.297..483S} showed that old stars in a certain range of effective temperatures display a similar amount of lithium ($\mathrm{A(Li^7)}\approx 2.2$), making up the so-called Spite plateau \citep[see also,][]{2010IAUS..268..201S}. \citet{2012ApJ...744..158C} used the measured number of baryons per photon from the Wilkinson
Microwave Anisotropy Probe satellite \citep{2011ApJS..192...18K} to constrain the primordial $\mathrm{A(Li^7)}\approx 2.72$ dex. Later on, \citet{2014JCAP...10..050C} updated their results by using the cosmological parameters determined by Planck \citep{2014A&A...571A..16P}, which resulted in a slightly lower average value, $\mathrm{A(Li^7)}\approx 2.69^{+0.04}_{-0.03}$ dex. The discrepancy between the measured $\mathrm{A(Li^7)}$ and its prediction from Big Bang Nucleosynthesis is known as the \say{cosmological lithium problem}.

As shown in the section above, applying a high-efficiency value of envelope overshooting during the early-MS might lead to an over-depletion of lithium in the Spite-plateau area. Indeed, \citet{2011MNRAS.414.1158C} investigated the convective envelope overshooting of the Sun by using helioseismic data. They found a mild value of $\Lambda_\mathrm{e}\approx 0.37$ reproduced very well the measured data. 
Later, \citet{2012MSAIS..22..233M} and \citet{2015MNRAS.452.3256F} proposed a new scenario to explain the $^7$Li depletion during the PMS. These authors suggested that a high-efficiency overshooting, mass-independent parameter, may cause a significant PMS $^7$Li destruction, but then, in the late PMS, a residual accretion can restore  $^7$Li almost to the pristine value.

In this section we take a further step in our model construction by introducing a simple scheme that is purely exploring the role of envelope overshooting. In particular, a milder value of the overshooting efficiency parameter is used during the evolution before it reaches the end of the MS. While for the post-MS phases the $\Lambda_\mathrm{e}=0.6$ is used base on the best-selected model on the CMD fitting and the onset's location of thermohaline mixing above.

\begin{figure}[t]
    \centering
    \includegraphics[width=\columnwidth]{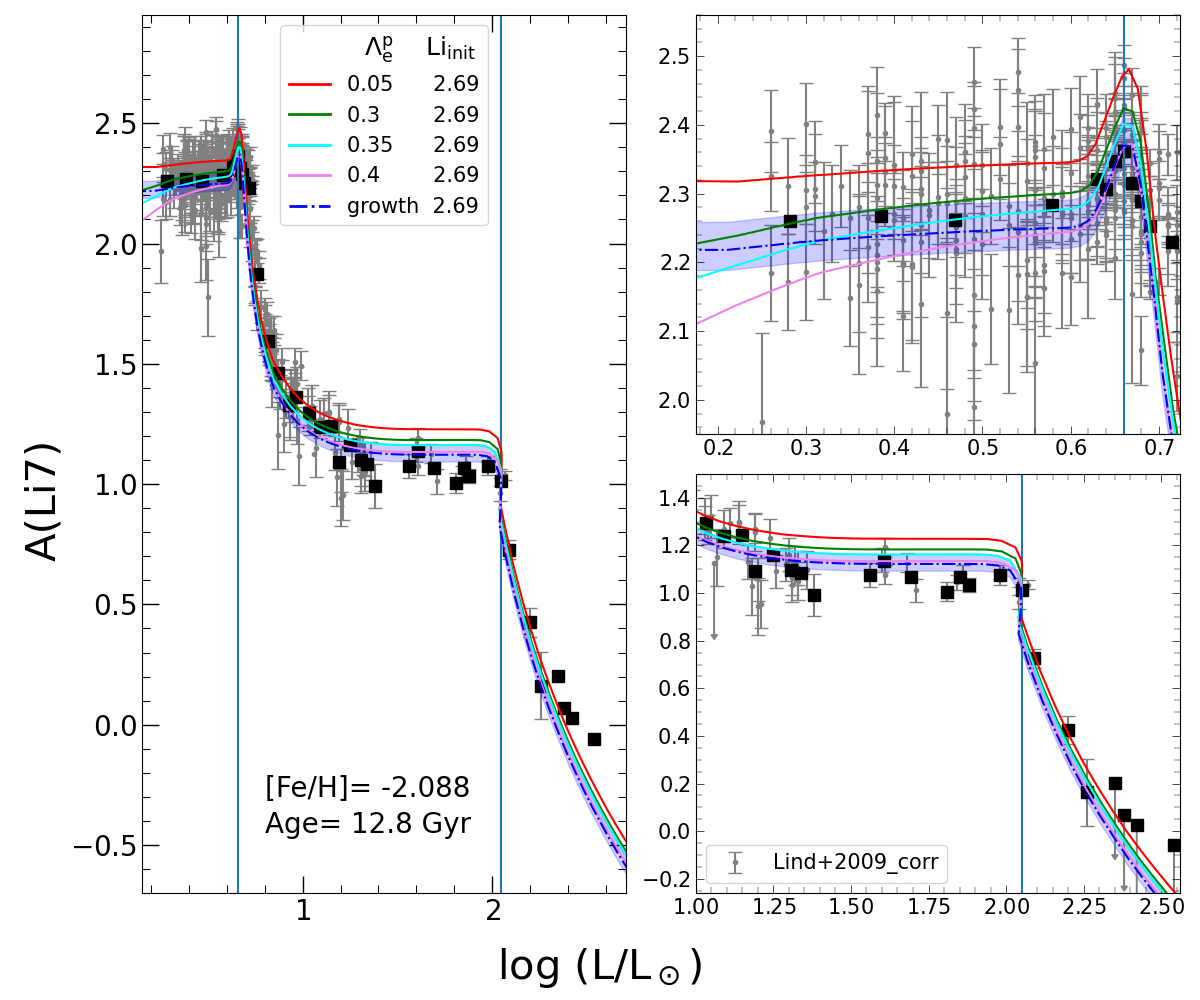}
    \caption{Variation of $^7$Li-abundance in average values (black squares) are superimposed with our models of various overshooting efficiencies at the early evolution. The applied values of $\Lambda_\mathrm{e}^\mathrm{p}$ and initial $\mathrm{A(Li^7)}$ are indicated in the label. The two panels on the right-hand side focus on the plateau (top-panel) and the RGB bump region (bottom-panel). With the re-derived 3D+NLTE data in grey dots. The shaded area marks the uncertainty coming from the primordial lithium, $\mathrm{A(^7Li)=2.69^{+0.04}_{-0.03}}$ of the "growth" model.}
    \label{ngc6397_Ali_iso_plt}
\end{figure}

We computed several grids of models with different values of envelope-overshoot efficiency for the early evolution (from the beginning of PMS to the end of MS) and applied the efficiency $\Lambda_\mathrm{e}=0.6$ for the rest of the evolution. To avoid confusion, the prior value that applied to early-MS phases is noted as $\Lambda_\mathrm{e}^\mathrm{p}$. The primordial lithium, $\mathrm{A(^7Li)}=2.69$, is adopted for the initial lithium abundance used in the models. For the sake of clarity regarding the trend of the observed abundances, especially in the region around the lithium peak on the subgiant branch, we use abundance values averaged over luminosity bins. 
Fig.~\ref{ngc6397_Ali_iso_plt} shows the averaged observed abundances and the predicted surface lithium abundances of models with different $\Lambda_\mathrm{e}^\mathrm{p}$. The constrained age and metallicity are adopted from the CMD fitting above. The averaged plateau clearly favours a model with a low value of envelope overshooting, $\Lambda_\mathrm{e}^\mathrm{p}=0.3-0.35$. However, the strength of the peak before the 1DU sets in identifies a best-fitting value, $\Lambda_\mathrm{e}^\mathrm{p}=0.4$. This result is in good agreement with the calibration of \citet{2011MNRAS.414.1158C}.

We take one more step on this aspect. As shown in Fig.~\ref{ngc6397_Ali_iso}, the colour bar shows a clear mass tendency of these stars, i.e., higher-mass stars have evolved to higher luminosities, while the Spite plateau shows an almost constant $\mathrm{A(^7Li)}$. The low-mass PMS stars have deep convective zones after they leave their birthlines, and the lower the mass, the deeper the convective boundary the star will have. Meanwhile, the temperature at the base of the envelope during the first few Myrs of the evolution is typically about $\sim 3\times 10^{6}$ K depends on its initial mass. Lithium, in turn, is very fragile as the nuclear reaction rate of $\mathrm{^7Li(p,\alpha)^4He}$ becomes efficient already in this range of temperatures. 
Therefore, the destruction of lithium during this early phase becomes very sensitive to the efficiency of overshooting, since the latter increases the temperature at the base of the mixed envelope. 
The variation of envelope overshooting efficiency on initial mass was already introduced in \citet{2012MNRAS.427..127B}, and recently \citet{2024ApJ...964...51A} required such dependency to reproduce the non-monotonic behaviour of the initial-final mass relation reported in \citet{2020NatAs...4.1102M}. We, in this work, apply this dependency on initial mass for $\Lambda_\mathrm{e}^\mathrm{p}$. Within the mass range from $0.64$-$0.86\,\Msun$ with step mass of $0.02\,\Msun$, the corresponding applied value of $\Lambda_\mathrm{e}^\mathrm{p}=0.05$-$0.60$ with a step of $0.05$. This model is shown by the dash-dotted blue line, labeled as \say{growth}, in Fig.~\ref{ngc6397_Ali_iso_plt}. It is evident that the new model reproduces the plateau, including the peak, quite well. However, we should emphasise that this new model is introduced to simply explore the impact of early-evolution variations of lithium before the onset of thermohaline mixing. While the effects of mixing processes to light elements such as lithium during the PMS and MS remains ambiguous \citep[see][and references therein]{2005ApJ...619..538R,2015MNRAS.452.3256F,2012A&A...539A..70E,2021A&A...646A..48D}.

Returning to the post-MS evolution, the averaged data display the renowned peak on the subgiant branch (at $\log L/L_\odot = 0.66$) observed in previous works \citep{2007ApJ...671..402K,2009A&A...503..545L} and understood as a short dredge-up of lithium from layers into which this element had previously settled by atomic diffusion.  
At the same time, the average luminosity of 3 stars at the RGB bump location is $\log L/L_\odot=2.05$. These two identifiable features are marked by the vertical lines in the figure panels. All our models demonstrate the fact that changing the overshooting efficiency doesn't change the location of the Li peak on the subgiant branch. However, changing the overshooting efficiency changes the location of the RGB bump. This is evidently seen in Fig.~\ref{ngc6397_Ali_iso_plt} and verified by Fig.~\ref{ngc6397_Ali_iso}, in which the peak's location of models with $\Lambda_\mathrm{e}=0.05$ to $\Lambda_\mathrm{e}=0.6$ show almost identical ($\leq 0.005$dex difference), while the RGB bump's location is about $\sim 0.15$dex difference. This suggests a robust way to calibrate the value of envelope overshooting efficiency by using the relative distance between the peak feature of $\mathrm{A(Li^7)}$ on the subgiant branch and the location of the RGB bump, where lithium is about to be destroyed due to the thermohaline mixing. This is a distance-independent method in the sense that the location of the RGB bump is measured relative to another feature of the same star cluster, assuring the robustness of this method. Indeed, the prediction of our model using the calibrated $\Lambda_\mathrm{e}=0.6$ shows the best fit to the data. Note that, in Fig.~\ref{ngc6397_Ali_iso_plt}, we apply a shift of $+0.05$ dex in $\log L$ to the data in order to match the peak with the isochrones. 
On this note, we notice that \citet{2009A&A...503..545L} uses a larger distance modulus, $12.57$ mag, which leads to an uncertainty of $\sim 0.2$ dex in $\log L$ when comparing with the value found by \citet{2018ApJ...864..147C}.

To conclude this subsection, we would like to emphasise that changing the efficiency of envelope overshooting at the early-MS phases does not change the photometry properties of the stars ($\log L$, $\log T_\mathrm{eff}$). Therefore, our model attempts for the early evolutionary phases do not change our conclusions above. Indeed, it shows sizeable effects on the variation of light elements such as lithium only during this evolution. Heavier elements such as $\mathrm{^{12}C}$ and $\mathrm{^{13}C}$ rather show no or very little effect in very low-mass models.

\begin{figure}[t]
    \centering
    \includegraphics[width=\columnwidth]{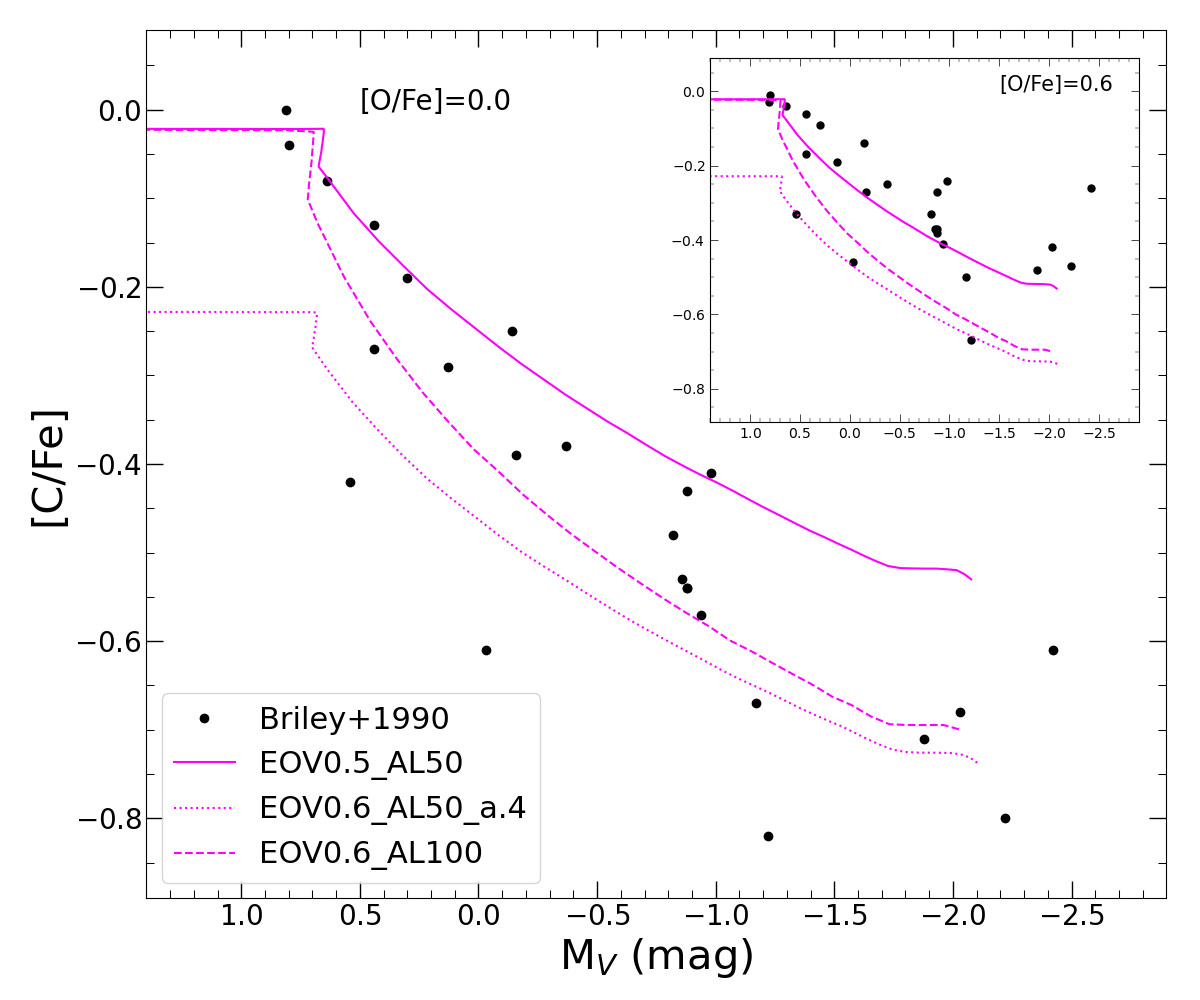}
    \caption{Variation of $^{12}$C abundance with absolute V-band magnitude. The best-fit isochrones in Fig.~\ref{ngc6397_cmd} are used in this plot together with the observed data from \citet{1990ApJ...359..307B} with two assumptions for oxygen enhancement ([O/Fe]=0.0 and 0.6).}
    \label{ngc6397_AC12_iso}
\end{figure}

\subsection{$^{12}$C abundances}
Besides lithium, carbon abundance is also commonly used to probe the internal mixing of RGB stars. As mentioned in Sect.~\ref{vari_3elements} the $^{12}$C/$^{13}$C ratio is a more reliable calibrator rather than the $^{12}$C abundance alone. However, due to the limitations of available isotopic data for carbon in NGC 6397, we can only carry on our comparison with the $^{12}$C abundances. Fig.~\ref{ngc6397_AC12_iso} shows A($^{12}$C) of 25 RGB stars in NGC 6397 by \citet{1990ApJ...359..307B} overplotted with the three models that we showed in Fig.~\ref{ngc6397_Ali_iso}.
To derive the absolute V-band magnitude, we apply the extinction coefficient from \citet{2018ApJ...864..147C}, namely, $A_V=0.76$ mag to our isochrones in Fig.~\ref{ngc6397_AC12_iso}.
As a matter of fact, the abundance data seems rather scattered, and our models with the values of thermohaline mixing efficiency $\alpha_\mathrm{th}=50$ to $100$ seemingly cover the spread of the data. At this point, we recall from Sect.~\ref{vari_3elements} (Fig.~\ref{chemi_single_mass_vari}) that the envelope overshooting does not play a significant role in the variation of carbon abundance.

The source for this spread in the observed data is not known to us. One of the hypotheses that could help us to comprehend this fact is the multiple populations in this cluster \citep[][and references therein]{2011A&A...527A.148L,2012ApJ...745...27M,2018MNRAS.475..257M}. To support this argument, the subplot in Fig.~\ref{ngc6397_AC12_iso} show the measured A($^{12}$C) by assuming an enhanced oxygen abundance, [O/Fe]=0.6, reported in \citet{1990ApJ...359..307B}. We see that our $\alpha$-enhanced model fits very well the three stars at the below-limit of the spread. However we should emphasise that a more detailed analysis would be needed to guide us to such a conclusion, and we reserve this topic for future works.

We also tested our models on the C abundance of another metal-poor, GC M92. We adopted the data from \citet{2001PASP..113..326B} and tried to fit them with our isochrones, $Z=0.0001$, $t=13$ Gyrs, $A_V=0.05$ mag. The result gave us an offset of about $\sim 0.2$\,dex higher than the data. We are not certain what causes this gap between the observed data and our isochrones, but we noticed that in their analysis, \citet{2001PASP..113..326B} reported different derived [C/Fe] values by using models with different input oxygen abundances. In other words, the input [O/Fe] apparently has a contribution to the uncertainty of the derived [C/Fe] values. For example, in many cases in their Table 4 there is a difference of about $0.1$\,dex between model using [O/Fe]=0.00 and 0.3. On this note, the recent analysis of globular clusters from the APOGEE survey with the BACCHUS code by \citet{2020MNRAS.492.1641M} shows the oxygen abundance of M92 is rather high $> 0.6$\,dex. Therefore, we reckon the inconsistency is due to the input [O/Fe] that was assumed to derive the C abundances in \citet{2001PASP..113..326B}. To conclude this section, we recognise that careful attention should be paid to the $\alpha$ elements in models when one uses C abundance to probe the internal mixing in low-mass RGB stars. Alternatively, the $^{12}$C/$^{13}$C ratio is more favourable, and we would like to explore this aspect in the near future.



\section{Distance-independent calibration of $\Lambda_\mathrm{e}$ from seismic data of GC M4}\label{EOV_calib}

It has been shown by \citet{2018ApJ...859..156K} that a distance independent calibration of the efficiency of the envelope overshooting can be obtained by analyzing the seismic properties of RGB bump stars.
Here we perform a similar analysis  using available seismic data of the globular cluster M4 provided by  \citet{2022MNRAS.515.3184H} (thereafter H22).

Firstly, during the evolution along the RGB phase, low-mass stars make a \say{zig-zag} path (or the RGB bump), and it is due to the interaction between the H-shell and the discontinuity. The encounter provides the H-shell more fuel and thus leads to a temporary decrease in luminosity, then shortly after, when the equilibrium is restored, the luminosity rises again. This implies that stars accumulate at this region of the colour-magnitude diagram (CMD), creating the RGB bump. Besides that, the location of the RGB bump is strongly correlated to the efficiency parameter of envelope overshooting. In particular, the more extended the envelope, i.e., the larger $\Lambda_\mathrm{e}$, the earlier the encounter occurs and hence the fainter the RGB bump. Comprehending these properties of the RGB bump, the star count distribution becomes a powerful tool to calibrate the efficiency parameter of the envelope overshooting.  For example, \citet{2018MNRAS.476..496F} used the star count distribution in a small bin of magnitude to calibrate the luminosity function of GC 47 Tuc and obtained a best-fitted value of $\Lambda_\mathrm{e}=0.5$. However, by using the magnitudes for such calibration, one has to cope with the uncertainty that comes from photometric properties such as distance modulus and dust extinction.

The seismic data sample provided in H22, that is claimed to be the largest seismic sample in a GC to date, contains seismic data for 59 red giant stars distributed below and above the location of the RGB bump. 
The observed seismic quantities, namely the frequency of maximum oscillation power ($\nu_\mathrm{max}$) and mean frequency separation of acoustic modes ($\Delta\nu$), can be expressed as  functions of intrinsic properties of the stars \citep[see e.g.][]{1986ApJ...306L..37U, 1995A&A...293...87K, 2012MNRAS.419.2077M} 
\begin{align}\label{numax_fomu}
    \frac{\nu_\mathrm{max}}{\nu_\mathrm{max,\odot}} =\left(\frac{M}{\Msun}\right)\left(\frac{R}{\Rsun}\right)^{-2}\left(\frac{\Teff}{T_{\mathrm{eff},\odot}}\right)^{-1/2},
\end{align}
\begin{align}\label{dnu_fomu}
    \frac{\Delta\nu}{\Delta\nu_\odot}=\left(\frac{M}{\Msun}\right)^{1/2}\left(\frac{R}{\Rsun}\right)^{-3/2}.
\end{align}
where  $M$, $R$ and $T_\mathrm{eff}$ are the stellar mass, radius, and effective temperature. 
In Eqs. \ref{numax_fomu} and \ref{dnu_fomu} the solar values are provided in H22,  $\nu_\mathrm{max,\odot}=3090 \mu$Hz, $\Delta\nu_\odot=135.1 \mu$Hz, $\Msun=1.989\times 10^{33}$ g, $\Rsun=6.9599\times 10^{10}$ cm, $T_\mathrm{eff,\odot}=5772$ K.

The comparison  between the observed seismic data of RGB bump stars with the corresponding ones predicted by models on the basis of the above relations can thus provide another check on the robustness of the envelope overshooting model.

Before proceeding with this test we recall that the predicted seismic properties of RGB bump stars change with both the metallicity and age.
To estimate these dependencies we have calculated theoretical isochrones for the metallicity and age ranges suitable for the GC M4. 
M4 is a moderately metal-poor globular cluster.  
In a recent analysis based on  high-resolution ESO/VLT FLAMES spectroscopy of 35 RGB stars spreading from the lower branch to above the bump region, \citet{2024MNRAS.52712120N} find a metallicity of $\feh=-1.13\pm 0.07$, in excellent agreement with previous works \citep[e.g.,][]{2011MNRAS.412...81M, 2012A&A...539A.157M, 2014AJ....147...25M, 2017A&A...607A.135W}. 
The cluster is confirmed to have $\alpha$-elements enhanced, e.g., \citet{2024MNRAS.52712120N} obtain $\mathrm{[O/Fe]} \sim 0.6$ dex, or \citet{2008A&A...490..625M} give $[\alpha/\mathrm{Fe}]=0.39\pm 0.05$ dex \citep[see also][]{2011A&A...535A..31V,2008ApJ...673..854Y}. 
The age of M4 varies between $11-13$ Gyr, e.g., \citet{2022A&A...662L...7T} estimate a range between $11-12$ Gyr, while \citet{2017A&A...607A.135W} give $11.50\pm 0.38$ Gyr which agrees very well with \citet{2015MNRAS.448..502M}, who find $11.81\pm 0.66$ Gyr;  \citet{2009ApJ...694.1498M} claim an age of  $12.65\pm 0.64$ Gyr while \citet{2002ApJ...574L.155H} derive an age of $12.7\pm 0.7$ Gyr.
We compute two initial metallicities $Z=0.001-0.002$, ($Y=0.250$, $0.252$ correspondingly), using the $\alpha$-enhanced mixtures \citep{2018MNRAS.476..496F}.
The isochrones are then obtained with the \textsc{trilegal} code for this purpose.

\begin{figure}[t]
    \centering    \includegraphics[width=\columnwidth]{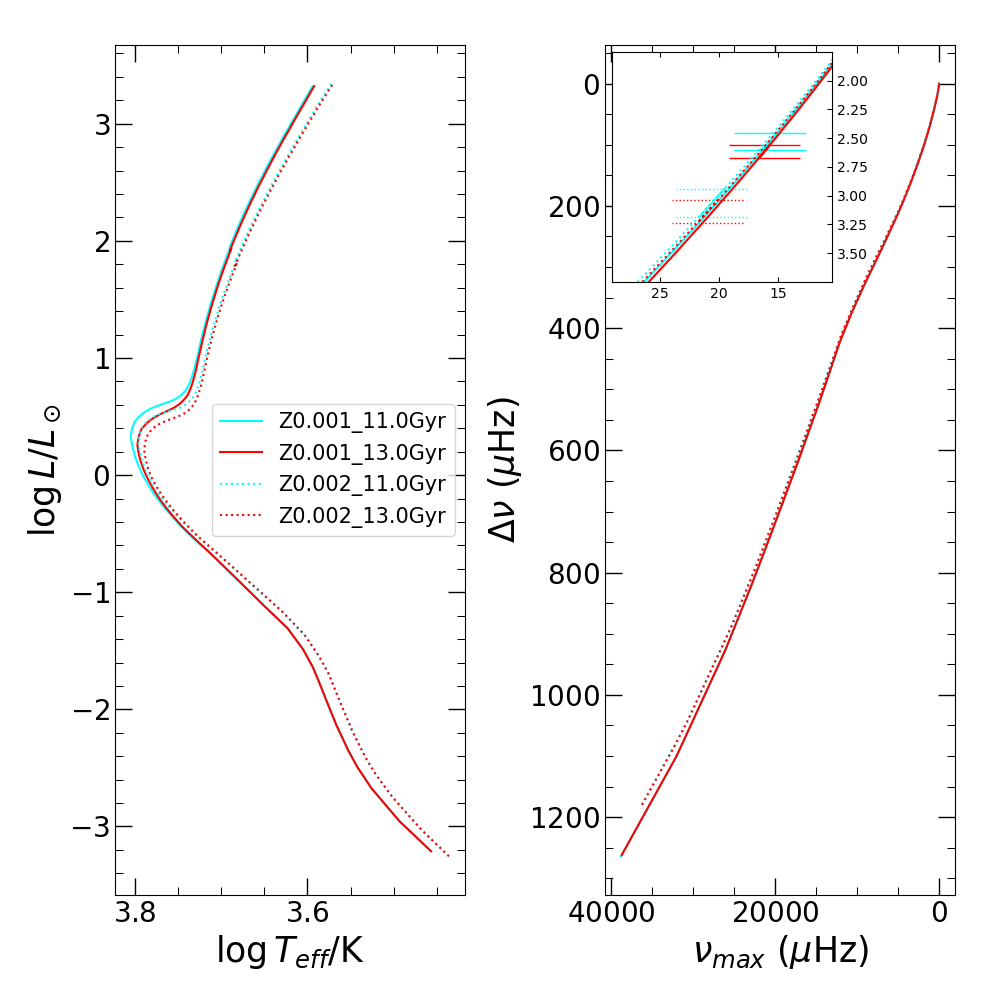}
    \caption{\textit{Left}: Hertzsprung–Russell diagram of four selected isochrones, with metallicity spanning from $0.001-0.002$ and age ranges from $11-13$ Gyr. \textit{Right}: the  seismic diagram corresponding to the HRD on the left side. The inset is a zoom-in of the region of the RGB bump. The horizontal bars mark the maximum and minimum of the bump.} 
    \label{seis_diagram}
\end{figure}

In Fig.~\ref{seis_diagram}, 
show the conversion from the theoretical isochrones ($\log L$, $\log T_\mathrm{eff}$) to the seismic diagram ($\Delta\nu$, $\nu_\mathrm{max}$), with the metallicity and age ranges suitable for the GC M4. 
Regarding to the location of the RGB bump (insert panel), we find that changing the age  from $11$ to $13$ Gyr affects $\Delta\nu$ by $\sim 0.1$ $\mu$Hz and  $\nu_\mathrm{max}$
by $\sim 0.5$ $\mu$Hz. Instead, by changing the metallicity from $Z=0.001$ and $0.002$, the variation of $\Delta\nu$ is $\sim 0.5$ $\mu$Hz and  
that of $\nu_\mathrm{max}$ is 
$\sim 4.5$ $\mu$Hz. 
Thus, to obtain a sound calibration from M4 data, it is more important to determine its precise metallicity than its age.

\begin{figure}[t]
    \centering
    \includegraphics[width=\columnwidth]{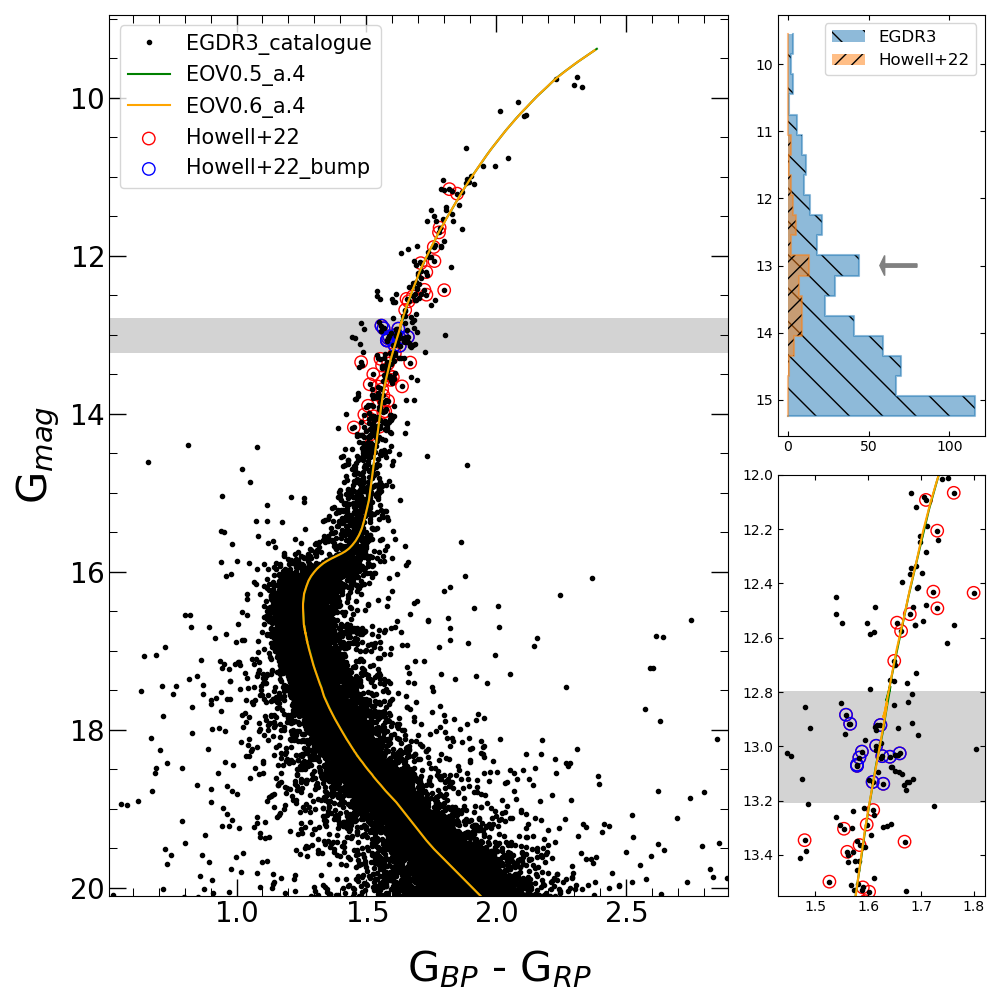}
    \caption{The CMD of M4 with data from the Gaia EDR3 catalogue \citet{2021MNRAS.505.5978V} with the isochrones shown in Fig.~\ref{m4_seismic_diagram}. The adopted parameters for isochrones are $12.6$ Gyr,  $Z=0.0015$, $(m-M)_0=11.35$ mag and $A_v=1.3$ mag. The open circles are stars from \citet{2022MNRAS.515.3184H} sample. \textit{Top-right}: star count histogram on $G_{mag}$ with a magnitude bin of $0.3$ dex along the RGB. \textit{Bottom-right}: zoom-in of the RGB bump area.}
    \label{M4_CMD}
\end{figure}

\begin{figure}[]
    \centering
\includegraphics[width=\columnwidth]{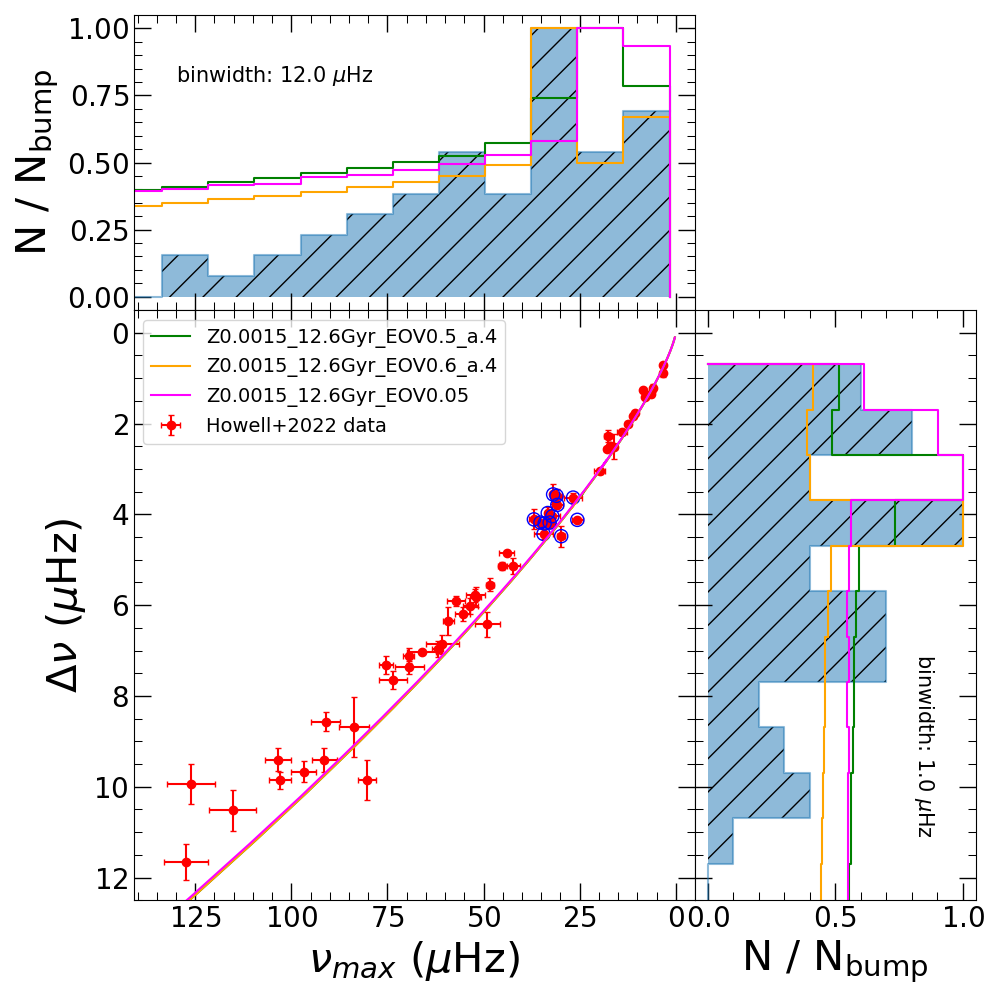}
    \caption{The seismic diagram of 58 RGB stars in \citet{2022MNRAS.515.3184H}. The selected isochrones with a variation of $\Lambda_\mathrm{e}=0.05,0.5, 0.6$ are superimposed by solid lines. The distinguished excess stars in the H22 sample are circled in blue. Star count histograms on $\nu_\mathrm{max}$ and $\Delta\nu$ are shown on the top and right panels, correspondingly, in which the blue-hatched area indicates the H22 data and the lines represent model predictions.}
    \label{m4_seismic_diagram}
\end{figure}

To perform the seismic calibration of the RGB bump of M4
we need to identify that the H22 sample has no significant selection bias on the evolutionary stage.
For this purpose we cross-match H22 stars with the cluster member catalogue of \citet{2021MNRAS.505.5978V} using Gaia EDR3 data. 
Fig.~\ref {M4_CMD}, shows the CMD of M4 in the Gaia EDR3/DR3 pass-bands, with data taken from \citet{2021MNRAS.505.5978V}. We have selected only the stars that have member probability larger $0.95$, and have eliminated the horizontal branch stars for the sake of cleanliness. 
The 58 RGB stars from H22 sample are displayed by the open circles. They are originally with Gaia DR2 photometry from H22, the Gaia DR3 information is obtained by cross-matching the Gaia DR2 neighbourhood catalogue (gaiaedr3.dr2\_neighbourhood), using Gaia DR2 source\_id provided by H22.
We highlight the likely RGB bump stars of H22 with blue open circles and the remaining stars in H22 with red open circles (star M4RGB169 is excluded because of the cross-matching). A zoom-in of this  RGB bump cross-match is shown in the bottom-right panel of Fig.~\ref {M4_CMD}. 
In the top-right panel of Fig.~\ref {M4_CMD}, we show the 
G-magnitude function of RGB stars from \citet{2021MNRAS.505.5978V} (blue histogram) together with that of H22 seismic data (orange histogram).
The magnitude of the RGB bump of the M4 Gaia member catalogue, $G_{mag}\sim 13$ mag,  is marked  by the grey-arrow. 
The G-magnitude function of the H22 seismic data (orange histogram) presents a peak that is consistent with the Gaia member RGB bump. This is a strong evidence that the H22 sample has no significant selection bias on the evolutionary stage compared to the Gaia \citet{2021MNRAS.505.5978V} sample. With this evidence, we can safely use the H22 seismic data to calibrate the RGB bump stellar models.

We now compare the observed seismic properties of H22 RGB stars and the ones predicted by assuming different efficiency of envelope overshooting, in Fig.~\ref{m4_seismic_diagram}.  
The top-left panel of Fig.~\ref{m4_seismic_diagram} shows the histogram of star count distribution in $\nu_\mathrm{max}$ with a step of $12$ $\mu$Hz. The bottom-right panel shows the $\Delta\nu$ histogram with a step of $1$ $\mu$Hz. 
The star count is normalised to its maximum value within the range, with the peak region referred to the distinguished RGB bump stars (blue-circle). 
The theoretical luminosity function is computed adopting an age of $12.6$ Gyr \citep[see][]{2009ApJ...694.1498M}, and an initial metallicity that produces the measured metallicity of \citet{2024MNRAS.52712120N}, in the bump region. It should be mentioned that \citet{2024MNRAS.52712120N} adopts \citet{2007SSRv..130..105G} solar compositions, while we adopt the compositions from \citet{2011SoPh..268..255C}. In other words, the rescaled metallicity due to the difference in solar reference composition is indeed $\feh=-1.2$ dex. This value corresponds to $Z=0.0015$ with our adopted $\mathrm{[\alpha/Fe]}=0.4$ mixture. 
The solid lines in Fig.~\ref{m4_seismic_diagram} show the prediction from theoretical models with this metallicity at varying the envelope overshooting parameter. The computed model with $\alpha$-enhanced mixture and $\Lambda_\mathrm{e}=0.6$ shows the best fit to the observed star count histogram, while the model with lower value $\Lambda_\mathrm{e}=0.05$ (adopted from \textsc{parsec} v1.2S) and $\Lambda_\mathrm{e}=0.5$ predict a too bright bump to reproduce the observed data.
It should be emphasised that this result holds for histograms performed in both $\Delta\nu$ and $\nu_\mathrm{max}$. We also check this best-fitted model on the CMD as shown in Fig.~\ref{M4_CMD}, together with the model of $\Lambda_\mathrm{e}=0.5$ for comparison. The isochrones are applied with 
the adopted distance modulus $(m-M)_0=11.35$ mag \citep[see][]{2021MNRAS.505.5957B}, and extinction coefficient $A_v=1.3$ mag \citep[see][and references therein]{2012AJ....144...25H}.

Furthermore, the multi-populations feature in the GC M4 has been studied in many works that showed that the He-content plays a key role. It's known that one population is formed with a He-enrichment, while the other population has a normal He-content that is compatible with the primordial value and locates at the redder part of the cluster \citep[e.g.,][]{2012ApJ...748...62V,2014ApJ...782...85V,2022ApJ...925..192D,2024arXiv240516505C}. However, the detailed analysis of the multiple populations in M4 is out of the scope of this paper. However, it's worth mentioning that we compute our models by following the standard enrichment law for He-content \citep{2012MNRAS.427..127B}. As a result, our isochrones give a good fit to the observed CMD, from the lower part of the MS up to the tip of the RGB. As an aside, we would like to mention that the He content can be the cause of the mismatch on the lower RGB in Fig.~\ref{m4_seismic_diagram}. However, this does not affect our conclusions since we focus on the upper part of the diagram where the RGB bump is located.

Finally, to conclude this section, we would like to stress that the model with $\Lambda_\mathrm{e}=0.6$ shows the best fit to the seismic's star count distribution of GC M4. This result supports our finding in Sect.~\ref{alpha_th_calib}. In other words, we have performed two independent calibrations, both distance- and reddening-independent, for the envelope overshooting parameter, and obtain good agreement with observations for $\Lambda_\mathrm{e}=0.6$. Although the two clusters we use for our calibrations have different metallicities, they are both in the low-metallicity domain [Fe/H]=[$-2.09$,$-1.21$], i.e.\ $\sim 0.9$\,dex apart. We note that the brightness of several GC's RGB bump across a wide range of metallicity is well reproduced with a single envelope overshooting parameter as stated in \citet{2018MNRAS.476..496F} (see their Fig. 14), in the range of [Fe/H]$> -1.3$ dex. Our results point in the same direction, but for a lower-metallicity domain.


\section{Summary and conclusions}\label{conclude}
We present in this paper, first, the implementation of thermohaline mixing in the \textsc{parsec} code, and second the interplay between thermohaline mixing and envelope overshooting in low-mass stars. We explore the effects of these mixing processes on the surface abundance of lithium and carbon in Sect.~\ref{vari_3elements}, together with variations in the adopted chemical mixtures. The efficiency parameters of these processes control the amount of abundances measurable at the stellar surface.
Using observed data on two well-studied globular clusters (NGC 6397 and M4), we perform the calibrations on the efficiency of the thermohaline mixing parameter, $\alpha_\mathrm{th}$, and on the envelope overshooting parameter, $\Lambda_\mathrm{e}$, in Sects.~ \ref{alpha_th_calib} and \ref{EOV_calib}. 

For theoretical models, we use the most recent version of the \textsc{parsec} code \citep{2019MNRAS.485.4641C,2022A&A...665A.126N} with the implemented thermohaline mixing. Four sets of initial metallicity, $Z=0.0001$, $0.0002$, $0.001$, $0.002$, are computed, with the variation of the efficiency parameters. Each metallicity set contains 14 stellar tracks with masses spanning from $0.64-0.94\Msun$. The corresponding isochrones are then produced by using the \textsc{trilegal} code with the adoption of very-low-mass models from previous \textsc{parsec v1.2s}. Then, they are complemented with the YBC bolometric correction tables for the benefit of our calibrations.

In this work, thermohaline mixing is treated as a diffusive process, and we follow the standard scheme of \citet{2007A&A...467L..15C,2010A&A...521A...9C} for modelling it. We carefully study the impact of thermohaline mixing on the evolution and the changes in chemical abundances.

We perform calibration on the thermohaline mixing efficiency parameter based on the observed $^7$Li abundances of stars in GC NGC 6397 from \citet{2009A&A...503..545L}. To work with the latest 3D+NLTE abundances, we re-derive the inferred abundances from the lines' equivalent widths by using the \textsc{breidablik} package \citep{2021MNRAS.500.2159W}. A relatively small value of 
$\alpha_\mathrm{th}=50$ is obtained to reproduce the lithium abundance of three stars above the RGB bump. 
Moreover, our attempted models with smaller envelope overshooting efficiencies for the PMS and MS phases, and $\mathrm{\Lambda_e}= 0.6$ for the post-MS, together with the thermohaline efficiency parameter $\alpha_\mathrm{th}=50$ gives the best fit to the 3D+NLTE lithium abundances of GC NGC 6397, from the Spite-plateau to the RGB stars. Furthermore, the luminosity difference between the Li-peak on the subgiant and the onset of thermohaline mixing at the RGB bump implies a robust calibration with $\mathrm{\Lambda_e}= 0.6$.

An independent and alternative approach to calibrate the efficiency of envelope overshooting is done by using the seismology data of GC M4. In this analysis, we focus on the star-count distribution in both the frequency of maximum oscillation power, $\nu_\mathrm{max}$, and the mean frequency separation of acoustic modes, $\Delta\nu$, to draw our conclusion on the best-fitting value for $\Lambda_\mathrm{e}$. As a result, we find that $\Lambda_\mathrm{e}=0.6$ is required to reproduce the accumulation of stars at the RGB bump of M4. This result agrees with the former one, and thus strengthen our conclusion.

Finally, we would like to stress that even though rotation has been implemented in the \textsc{parsec} code \citep[][]{2019MNRAS.485.4641C,2022A&A...665A.126N}, we do not consider it in the present work. We focus on the effect of convective envelope overshooting and the thermohaline mixing along the post-MS phases. The impact of rotation on thermohaline mixing has already been investigated in the literature \citep[e.g.,][]{2010A&A...522A..10C,2011A&A...536A..28L}. Moreover, \citet{2013A&A...553A...1M} pointed out that the horizontal turbulence in rotating stars can inhibit thermohaline mixing, which led to an even more complex scenario. A project combining all these effects is deferred to the future.

The re-derived 3D+NLTE lithium abundance of NGC 6397, together with the stellar tracks of our best-fitted models that are presented in this paper can be achieved at a dedicated ZENODO dataset\footnote{\url{https://doi.org/10.5281/zenodo.13283724}}. The isochrones with different photometric pass-bands can be provided upon request.

\begin{acknowledgements}
This project has received funding from the European Union’s Horizon 2020 research and innovation programme under grant agreement No 101008324 (ChETEC-INFRA). CTN also thanks Madeline Howell for providing useful seismic data information.
\end{acknowledgements}


\bibliographystyle{aa}
\bibliography{references} 

\makeatletter
\def\thebiblio#1{%
 \list{}{\usecounter{dummy}%
         \labelwidth\z@
         \leftmargin 1.5em
         \itemsep \z@
         \itemindent-\leftmargin}
 \reset@font\small
 \parindent\z@
 \parskip\z@ plus .1pt\relax
 \def\newblock{\hskip .11em plus .33em minus .07em}
 \sloppy\clubpenalty4000\widowpenalty4000
 \sfcode`\.=1000\relax
}
\let\endthebiblio=\endlist
\makeatother

\label{lastpage}

\end{document}